\documentclass[%
 onecolumn,
superscriptaddress,
 amsmath,amssymb,
 aps,
longbibliography
]{revtex4-2}

\usepackage{graphicx}% Include figure files
\usepackage{dcolumn}% Align table columns on decimal point
\usepackage{bm}% bold math
\usepackage[mathlines]{lineno}% Enable numbering of text and display math
\usepackage[colorlinks,citecolor=blue]{hyperref}
\usepackage[capitalise,nameinlink]{cleveref}
\hypersetup{colorlinks=true, linkcolor=blue}
\usepackage{subfigure}
\usepackage{soul}
\usepackage{float}

\renewcommand{\thetable}{\arabic{table}}%
\renewcommand{\thefigure}{\arabic{figure}}%

\begin{document}

\preprint{APS/123-QED}

\title{Hydrodynamics of a single filament moving in a fluid spherical membrane}% Force line breaks with \\

\author{Wenzheng Shi}
%\thanks{Contributed equally}
\affiliation{Department of Applied Physical Sciences, The University of North Carolina at Chapel Hill, Chapel Hill, North Carolina 27599, USA}

\author{Moslem Moradi} 
\affiliation{Department of Applied Physical Sciences, The University of North Carolina at Chapel Hill, Chapel Hill, North Carolina 27599, USA}

\author{Ehssan Nazockdast} 
\email{ehssan@email.unc.edu}
\affiliation{Department of Applied Physical Sciences, The University of North Carolina at Chapel Hill, Chapel Hill, North Carolina 27599, USA}

\date{\today}% It is always \today, today,

\begin{abstract}
Dynamic organization of the cytoskeletal filaments and rod-like proteins in the cell membrane and other biological interfaces occurs in many cellular processes, including cell division, membrane transport and morphogenesis. The filaments dynamics is determined, in part, by their membrane-mediated hydrodynamic interactions. Previous modeling studies have considered the dynamics of a single rod on fluid planar membranes. We extend these studies to the more physiologically relevant case of a single filament moving in a \emph{spherical} membrane. 
Specifically, we use a slender-body formulation to compute the translational and rotational resistance of a single filament of length $L$ moving in a membrane of radius $R$ and 2D viscosity $\eta_m$, and surrounded on its interior and exterior with Newtonian fluids of viscosities $\eta^{-}$ and $\eta^{+}$. 
We first discuss the case 
where the filament's curvature is at its minimum $\kappa=1/R$. 
We show that the boundedness of spherical geometry gives rise to flow confinement effects that increase in strength with increasing the ratio of filament's length to membrane radius $L/R$. 
These confinement flows only result in a mild increase in  
filament's resistance along its axis, $\xi_{\parallel}$, and its rotational resistance, $\xi_{\Omega}$. As a result, our predictions of $\xi_\parallel$ and $\xi_{\Omega}$ can be quantitatively mapped to the results on a planar membrane, when the momentum transfer length-scale is modified from $\ell_0=(\eta^{+}+\eta^{-})/\eta_{m}$ in planar membranes to  $\ell^\star=(\ell_0^{-1}+R^{-1})^{-1}$. 
In contrast, we find that the drag in perpendicular direction, $\xi_\perp$, increases superlinearly with the filament's length, when $L/R >1$ and ultimately $\xi_\perp \to \infty$ as $L/R \to \pi$.
Next, we consider the effect of the filament's curvature, $\kappa$, on its parallel motion, while fixing the membrane's radius. We show that the flow around the filament becomes increasingly more asymmetric with increasing its curvature. These flow asymmetries induce a net torque on the filament, coupling its parallel and rotational dynamics. 
This coupling becomes stronger with increasing $L/R$ and $\kappa$. 

\begin{description}
\item[PACS numbers]
May be entered using the \verb+\pacs{#1}+ command.
\end{description}
\end{abstract}

\pacs{Valid PACS appear here}% PACS, the Physics and Astronomy
\maketitle

\section{\label{sec:Intro}Introduction}

Many cellular processes involve the transport of rod-like proteins and biopolymers on curved fluid-like interfaces. One example is the continuous reorganization of the cell cortex made of dynamic network of actin filaments \citep{Rizzelli2020}, which determines cell's mechanical integrity and drives cell division. Another example is the transport and organization of rod-like proteins that preferentially bind to areas of specific curvature, such as septin oligomers \citep{Cannon2019} and BAR domains \citep{simunovic2013linear}. One required ingredient for simulating the dynamics of filaments in fluid membranes is the hydrodynamic resistance/mobility of a single filament.
Moreover, mobility/resistance functions are typically needed to measure the interfacial rheology of the membrane from 
particles' motion in passive and active microrheology \citep{kusumi2005paradigm, prasad2006two, verwijlen2011study, zell2014surface}. \par
The first theoretical studies of the mobility of a membrane-bound particle goes back to the works of  \citet{Saffman1975} and \citet{Saffman1976} who calculated the mobility of a thin disk of radius $a$, bound to a planar membrane; the membrane is modeled as a 2D Newtonian fluid of viscosity $\eta_m$, and overlaying an infinitely bound Newtonian fluid of viscosity $\eta$. The mobility takes the form: $\chi=(1/4\pi \eta_m) \left[\ln (2\ell_0/a)-\gamma\right]$, where $\gamma \approx 0.5772$ is Euler's constant and $\ell_0=\eta_m/\eta \gg a$ is the length-scale over which the momentum is transported from the membrane to the overlaying 3D fluid domain, referred to hereafter as Saffman-Delbr\"{u}ck (SD) length. Coupling the membrane and the overlaying fluid flows introduces SD length-scale and, hence, removes the well-known Stokes paradox associated with solutions of 2D Stokes problems in free space. Note that for a particle significantly smaller than SD length, its dimensions affect the mobility through the weak logarithmic term $\ln (2\ell_0/a)$.
Some experimental studies on the dynamics of membrane-bound proteins \citep{Ramadurai2009, Hormel2014} have found good agreement between Saffman's predictions and experiments, while others \citep{gambin2006lateral} found the mobility to scale inversely with particle size. These disagreements were
consequently described by models that account for the additional flow dissipation due to local deformation of the membrane induced by interactions with the bound particle \citep{Naji2007}, and membrane structure and mechanics \citep{Quemeneur2014, sigurdsson2013hybrid, camley2012contributions}. 
\par
\citet{Evans1988} and \citet{Stone1998} considered the case of a disk bound to a planar membrane overlaying a 3D fluid domain of finite depth. These studies show that the momentum transfer from the membrane to 3D fluid domain occurs at length-scale $\ell_H= \sqrt{\ell_0 H}$, where $H$ is the 3D fluid depth. 
\citet{Oppenheimer2009} studied the hydrodynamic interactions (HIs) between two Brownian disks in the same system.
\citet{oppenheimer2019rotating} extended this analysis to a suspension of rotors, and showed that the suspension can crystallize by a combination of hydrodynamic and steric interactions between the particles, while ignoring any of these interactions does not lead to crystallization. 
\citet{manikantan2020tunable} numerically studied particles with a force dipole acting on them, as a model for membrane-bound active particles. Their results show that large-scale particle aggregation can be controlled through tuning the depth of the surrounding 3D fluid domain.
 %They show that the confinement introduces another hydrodynamic screening length scale $\ell^\star=\sqrt{H\ell_0}$.
%, and lateral diffusion of particle among multiple immobile particles \citep{Dodd1995, Singh2019}. 
\par
\citet{Levine2004} considered the mobility of a rod-like particle moving with a constant transnational and rotational velocity in a planar membrane. The authors used a slender-body formulation and the fundamental solutions to a point-force on a planar membrane (planar Green's function) to describe the disturbance flows induced by the filament's motion. 
They found that for $L/\ell_0 \ll 1$, the drag coefficient is independent of rod orientation and asymptotes to SD results for a disk of radius $L$. 
%This is to be expected as the dynamics is determined by membrane viscosity and independent of 3D fluid viscosities. 
When $L/\ell_0\gg 1$, the drag coefficient in parallel direction retains the logarithmic correction that is observed in 3D flows: $\xi_{\parallel}/(4\pi\eta_{m})\approx 0.25 \left(L/\ell_{0}\right)\left(\ln(0.43L/\ell_0)\right)^{-1}$, with the difference that the effective filament radius is now $\ell_0$, rather than the physical radius $a$. In contrast, the drag coefficient in the perpendicular direction becomes purely linear in length, $\xi_{\perp}/(4\pi\eta_{m})\approx 0.25 \left(L/\ell_{0}\right)$, and significantly larger than the parallel drag. In comparison the ratio of perpendicular to parallel drag of filaments in 3D flow is nearly two. 
These predictions were corroborated with microrheological measurements of the translational and rotational drags over the range of $0.01 \le L/\ell_0 \le 10$ \citep{Klopp2017}. 
\par
This qualitative difference between a 2D membrane and a 3D fluid domain can be understood by noting that the planar Green's function along the direction of the applied force and perpendicular to it, when $L/\ell_{0}\gg1$, decay as $G_\parallel \sim {1}/{\tilde{r}}$ and $G_\perp\sim {1}/{\tilde{r}^2}$, respectively, where $\tilde{r}={r}/{\ell_0}$ is the dimensionless distance from the applied force (in comparison, the 3D Green's function decays as ${1}/{r}$ in both directions). Thus, when $L/\ell_0\gg 1$, the filament's mobility scales with $\chi_{(\parallel,\perp)}\sim (\ell_0/L) \int_1^{L/\ell_0} G_{(\parallel,\perp)} (\tilde{r}) d \tilde{r}$. 
This leads to a logarithmic contribution in the parallel direction and
a constant in the perpendicular direction (free draining limit), when $L/\ell_0 \gg 1$ \citep{manikantan2020surfactant}.
\par
\citet{fischer2004drag} and \citet{verwijlen2011study}
extended the work of \citet{Levine2004} to study the dynamics of rod-like inclusions in membranes overlaying a fluid domain of finite depth, and explored the consequences of this finite depth on the accuracy of interfacial rheological measurements.
\par
Membranes are typically curved in biological and synthetic applications. Membrane's curvature and topology produces novel features in the interface-bound transport processes that make it distinct from a planar membrane \citep{scriven1960dynamics, secomb1982surface, mavrovouniotis1993_1, mavrovouniotis1993_2, woodhouse2012shear, honerkamp2013membrane}. Different statistical mechanical aspects of curved membranes are surveyed in \cite{nelson2004statistical}.
%. Early formulation of conservation laws on curved interfaces include the pioneering work of \cite{scriven1960dynamics}, detailed derivations of \cite{aris2012vectors}, followed by \cite{secomb1982surface}, \cite{mavrovouniotis1993_1} and \cite{mavrovouniotis1993_2}. 
 %The equations describing the flow in the membrane and the adjacent fluid are complex for general curved geometries, and they are typically solved numerically;  see the recent work by \cite{torres2019modelling}, \cite{ tozzi2019out}, and \cite{pozrikidis2001interfacial} for a review on continuum simulations of flows in deformable membranes with different constitutive equations.  
%\par
Most studies on the dynamics of inclusions bound to the membrane are done on planar membranes, and  the number of studies on curved membranes are in comparison small \citep{sigurdsson2016hydrodynamic, rower2019surface}. In the special cases of spherical and cylindrical membranes the Gaussian curvature remains constant over the surface. This greatly simplifies the equations, allowing for analytical calculations of fundamental solutions to a point-force, a point-torque, and a torque-dipole \citep{Henle2008, Henle2010, samanta2021vortex}. These fundamental solutions can be used to describe the flow disturbances due to the presence of other bodies in the membrane, which is similar to the singular methods for 3D Stokes flow suspensions \citep{happel2012low, Kim2013}. This is also the methodology used in this study. 
\par
\citet{Henle2010} studied the translational mobility of a disk-like inclusion of radius $a$ on \emph{non-deformable} spherical and cylindrical membranes of radius $R$. In the limit of small curvature $R/\ell_0 \gg 1$, one recovers the planar membrane results. At larger curvatures $R/\ell_0 \ll 1$, the mobility is reduced with the logarithmic term now being defined as $\ln(R/a)$ in contrast to $\ln (\ell_0/a)$ on planar membranes. In a recent study, \citet{samanta2021vortex} used the same framework to study the effect of the membrane curvature on the hydrodynamic interactions (HIs) and the ensued nonlinear dynamics of rotors bound to spherical membranes. \citet{bagaria2021dynamics} investigated the curvature confinement effect on the aggregation of force dipoles on spherical membrane. 
\par
The current study extends the work of \citet{Levine2004} to filaments moving in spherical membranes. Specifically, we consider the case of a filament of length $L$ and constant curvature, $\kappa$, along its axis, moving in a membrane of radius $R$ with 2D viscosity $\eta_m$, and surrounded by the interior and exterior Newtonian fluids of viscosities $\eta^{-}$ and $\eta^{+}$, where the SD length is redefined as $\ell_{0}=\eta_{m}/(\eta^{+}+\eta^{-})$. In \S\ref{sec:resist} we study the filament's drag when its curvature is at its minimum $\kappa_{\text{min}}=1/R$. This is the filament's likely conformation, when the characteristic bending forces are comparable to filament-membrane interaction forces in the radial direction and both are significantly larger than the thermal forces. 
We use a slender-body formulation to compute the filament's drag coefficients along its axis, $\xi_{\parallel}$, and perpendicular to it, $\xi_{\perp}$, as well as its rotational drag coefficient, $\xi_{\Omega}$ in a wide range of $\ell_0/R$, and $L/R$. 
\par
When membrane radius is much larger than SD length, $\ell_{0}/R\ll 1$, the filament's dynamics becomes independent of the membrane radius and our results approach those of a planar membrane given by \citet{Levine2004}. 
In the opposite limit of $\ell_{0}/R\gg 1$, the momentum transfer to the overlaying fluid domain occurs over membrane's radius $R$, and the dynamics become independent of $\ell_{0}$. 
Here, we find that the transnational resistance in the direction parallel to the filament and the rotational resistance closely follow the results of a planar membrane, if SD length is replaced with the membrane radius. In contrast, the resistance in perpendicular direction, $\xi_\perp$, shows strong positive deviations from the planar membrane results. We show that these deviations arise due to flow confinement effects on a spherical membrane, which has no analog in free 2D planar membranes. As a result, $\xi_\perp$ strongly increases with the filament's length when $L/R\sim\mathcal{O}(1)$, and $\xi_\perp \to \infty$ as $L/R\to\pi$. 
\par
In \S\ref{sec:off-equator} we study the effect of the filament's curvature on its dynamics. The motivation for studying this problem is to model 
the dynamics of rod-like biopolymers with an intrinsic (preferred) curvature, which preferentially bind to areas of the membrane that match their curvature. One example includes the crescent-shaped BAR domains that play a key role in inducing and sensing membrane curvature in cellular processes \citep{simunovic2013linear}. 
We present the filament with curvature $\kappa$ by aligning it along the azimuthal direction of the spherical membrane and placing it at a distance $h= \sqrt{R^2-\kappa^{-2}}$ from the equator in the $z$ direction. 
We show that filament's motion along its axis produces asymmetric flows, resulting in a net torque on the filament. This coupling between parallel and rotational motion is increased with filament's length and curvature. 

\section{Formulation} \label{sec:Formula}
We consider the motion of a filament with length $L$ and \emph{constant curvature} along its length, bound to a fixed (undeformable) spherical membrane of radius $R$ and Newtonian viscosity $\eta_m$. 
Due to spherical symmetry, any configuration of a filament with constant curvature can be mapped into the filament being aligned along the azimuthal direction and with a distance $h$ along the $z-$axis, which is shown in \cref{fig:illustration}.
The filament's curvature is at its minimum, $\kappa_\text{min}=1/R$, when it is placed on the equator ($h=0$). This is the most likely conformation of the filament if the filament does not have an intrinsic curvature and the characteristic bending/elastic forces are significantly larger than thermal forces. 
We begin by studying the dynamics in this limit  in \S\ref{sec:resist}. Due to geometric and flow symmetries the translational and rotational modes of motion of the filament are decoupled.  Thus, the translational resistance tensor can be generally defined as $\boldsymbol{\xi}=\xi_\parallel \mathbf{q}\mathbf{q}+ \xi_\perp\left(\mathbf{I}-\mathbf{q}\mathbf{q}\right)$, where $\mathbf{I}$ is the identity matrix and $\mathbf{q}$ is the unit vector along the filament main axis. 
\par
As shown in \cref{fig:illustration}, the general case of a filament with curvature $\kappa > 1/R$ is equivalent to placing the filament at 
$h=\pm \sqrt{R^2-\kappa^{-2}}$ away from the equator and aligned along the azimuthal direction $\phi$, which corresponds to constant polar angle $\theta=\arcsin (\kappa R)^{-1}$. Moving the filament along its axis produced asymmetric flows, and a net torque on the filament, resulting in the coupling of the filament's parallel and rotational motion. This is discussed in \S \ref{sec:off-equator}.  
\par
We use a slender-body formulation to model the flow disturbances induced by the filament with a distribution of force densities, $\mathbf{f}(s)$, along the filament's length:
\begin{equation}
    \mathbf{u}(\mathbf{x}_m)=\int_{-L/2}^{L/2}
    \mathbf{G}(\mathbf{x}_m,\mathbf{x}^0_m(s))
    \cdot\mathbf{f}(s)\mathrm{d}s,
    \label{eq:sbt1}
\end{equation}
where $s \in [-L/2,L/2]$ denotes the filament's arclength, $\mathbf{x}_m$ and $\mathbf{x}^0_m$ are points on the membrane surface, and $\mathbf{G}(\mathbf{x}_m,\mathbf{x}^0_m(s))$
is the Green’s function of membrane-3D fluids coupled system in response to a point-force applied on the membrane at position $\mathbf{x}^0_m$. 
Note that the Green's function scales as $\ln |s-s^\prime|$ when $s-s^\prime\to 0$, and thus diverges as $|s-s^\prime|\to 0$. But,  our resolution studies presented in the supplementary materials show that the integrals are numerically convergent and no further regularization is needed.
\par
Assuming flow incompressibility on the membrane surface and the adjacent 3D flows and negligible inertia, the associated momentum and continuity equation for the membrane and 3D fluid domains are \citep{samanta2021vortex, Henle2010}: 
\begin{subequations} 
\begin{align}
\label{eq:stokes_3D}
&\text{3D fluids:}& &\eta^{\pm}\nabla^{2}\mathbf{u}^{\pm}-\nabla p^{\pm}=\mathbf{0},&
\nabla\cdot\mathbf{u}^{\pm}=0,&\\
\label{eq:stokes_sphere}
&\text{Membrane:}& &\eta_{m}\left(\Delta_\gamma \mathbf{u}_{m}
+K(\mathbf{x}_m)\mathbf{u}_{m}\right)-\nabla_\gamma p_{m} +\mathbf{T}|_{r=R}=\mathbf{0}, &\nabla_\gamma\cdot \mathbf{u}_{m}=0, &
\end{align}
\label{eq:Eqs}
\end{subequations}
where $\mathbf{u}^\pm$ and $p^\pm$ are the velocity and pressure fields in 3D fluid domains, and $\mathbf{u}_m$ and $p_m$ are the velocity and pressure fields in the membrane;  $\Delta_\gamma$ and $\nabla_{\gamma}\cdot$ are the surface (defined by $\gamma$) Laplacian and Divergence operators,  $K(\mathbf{x}_m)$ is the local Gaussian curvature of the surface, and $\mathbf{T}|_{r=R}=\left(\boldsymbol{\sigma}^{+}(\mathbf{x}_m)-\boldsymbol{\sigma}^{-}(\mathbf{x}_m)\right)\cdot \mathbf{n}(\mathbf{x}_m)$ is the traction applied from the surrounding 3D fluid domains on the membrane, where $\boldsymbol{\sigma}^\pm$ denotes the 3D fluid stress and $\mathbf{n}(\mathbf{x}_m)$ is the surface normal vector pointing towards the exterior domain.  
\par
The coupling between the membrane and 3D flows is enforced by the continuity of forces and velocities across the membrane surface. The force continuity is already imposed through including the traction term, $\mathbf{T}|_{r=R}$, in \cref{eq:stokes_sphere}. The radial velocity is zero at $r=R$, for a fixed spherical membrane, which results in $u_r^\pm=0$ throughout both 3D fluid domains. Other boundary conditions (BCs) for 3D flows include velocities decaying to zero on the exterior domain, $\lim_{r\to \infty} u_{\theta,\phi}^+ (r)\to 0$. 
Finally, requiring the velocity and stress fields to be finite at the interior domain's center, $r=0$, provides the sufficient BCs.
\par
The Gaussian curvature  of a sphere is a constant, $K(x)=R^{-2}$, which greatly simplifies \cref{eq:Eqs}, allowing one to obtain analytical solutions for Green's function in terms of the position of the applied force $(\theta_0,\phi_0)$ and an arbitrary target point on the sphere $(\theta,\phi)$. Here, $\theta \in [0,\pi]$ and $\phi \in (0,2\pi]$ are the polar and azimuthal angles in spherical coordinate. The detailed derivation of the Green's function is outlined in \cite{Henle2010} and \cite{samanta2021vortex}. The expressions for all components of the Green's function ($G_{\theta\theta},\, G_{\theta\phi},\, G_{\phi\theta},\, G_{\phi\phi}$) are provided in \cref{sec:appA} for completeness. 
\par
To calculate the resistance in parallel ($\xi_\parallel$) and perpendicular ($\xi_\perp$) directions, we set the filament velocity as a constant in each direction and compute for the distribution of force density on the filament by solving the following integral equation: 
\begin{equation} 
\label{eq:FiberMobility}
\mathbf{U}(s)= \int_{-L/2}^{L/2}\mathbf{G}
(\mathbf{X}(s)-\boldsymbol{X}(s^\prime))\cdot  \mathbf{f}(s^\prime)\mathrm{d}s^\prime,
\end{equation}
where $\mathbf{X}(s)$ is a point located at $s$ arclength of the filament. 
Equation \ref{eq:FiberMobility} is a Fredholm integral equation of first kind. This class of integral equations are known to be ill-posed leading to a loss of convergence and numerical inaccuracy.
To circumvent this issue we use a regularization technique, analogous to 
those used in non-local slender-body theories of filaments moving in 3D fluid domains \citep{Tornberg2004, Nazockdast2017}, to 
transform the equation to a second-kind Fredholm integral equation, which can, then, be accurately solved using different numerical integration techniques. 
\par
We divide the domain of integration into a short local domain around $s$, $\Upsilon_\text{L} (s^\prime)= [s-\frac{\delta L}{2},\, s+\frac{\delta L}{2}]$ and  the remainder of the filament length, $\Upsilon_\text{NL} ( s^\prime )= [-\frac{L}{2},s-\frac{\delta{L}}{2})\cup(s+\frac{\delta{L}}{2},\frac{L}{2}]$, where $\frac{\delta L}{L}\ll 1$, $\frac{\delta L}{R} \ll 1$ and "L" and "NL" correspond to the local and nonlocal domains \citep{lighthill1976flagellar}. 
The local domain near the ends of the filament modifies to 
$\Upsilon_\text{L}(s^\prime ) = [\frac{L}{2}-\delta L,\,\frac{L}{2}]$ and $\Upsilon_\text{L}(s^\prime ) = [-\frac{L}{2},\,-\frac{L}{2}+\delta L]$. 
Next, we assume the force density remains uniform within the local domain: $\mathbf{f} (s^\prime)\approx \mathbf{f}(s)$. The modified integral equation is   
\begin{equation}
\label{eq:FiberMobility2}
\mathbf{U}(s)= 
%\left(\int_{-\frac{L}{2}}^{s-\frac{\delta{L}}{2}}+\int_{s+\frac{\delta{L}}{2}}^{\frac{L}{2}} \right)
\int_{\Upsilon_\text{NL}}
\mathbf{G}(\mathbf{X}(s)-\boldsymbol{X}(s^\prime))\cdot  \mathbf{f}(s^\prime)\mathrm{d}s^\prime
+ \left(\int_{\Upsilon_\text{L}} \mathbf{G}(\mathbf{X}(s)-\mathbf{X}(s^\prime))\mathrm{d}s^\prime \right)\cdot \mathbf{f}(s). 
\end{equation}
\par
The second term on the right hand side makes \cref{eq:FiberMobility2} a second kind Fredholm integral equation, which can be solved by 
collocation methods described briefly in \S\ref{sec:num_method}.
The computed force density is, then, integrated to obtain the total force on the rod. The resistance is determined by taking the ratio of this force to velocity in $\parallel$ and $\perp$ directions.
\par
The rotational resistance can also be computed using the same formulation with the difference that the set velocity correspond to a pure rotational motion: $\mathbf{U}_{\Omega}(\mathbf{X}(s))=\Omega s \mathbf{q}_\perp$, where $\Omega$ is the angular rotation, and $\mathbf{q}_\perp$ is the filament's transverse direction. The rotational resistance is the ratio of the torque to angular rotation.

\section{Numerical implementation} \label{sec:num_method}
\noindent
We solve for the force distribution in \cref{eq:FiberMobility2}, by discretizing the rod into $N$ equally spaced points, and use trapezoidal integration. This leads to $N$ linear system of equations in the matrix form: $\mathcal{U}=\mathcal{A}\cdot \mathcal{F}$, where $\mathcal{U}$ is the $N\times 1$ array representing unit velocities along the rod at a given direction, $\mathcal{A}$ is the $N\times N$ matrix representing the HIs, and $\mathcal{F}$ is the unknown forces along the rod. 
We solve this system using direct solution methods. 
The computed forces are, then, integrated to give the resistance in parallel and perpendicular directions. 
\par
The length of the local domain is set to $\frac{\delta L}{ L}=\frac{1}{30}$, for all the results presented here. Varying this length in the range $10^{-2} \le \frac{\delta L}{L}\le 10^{-1}$ changed the results by less than $5\%$. Resolution studies at a given $\delta L$ show that our method is second-order accurate. These results are provided in supplementary materials. The reported results were obtained using $N=1021$ discretization points. 
The computations can be made far more efficient requiring less number of points. But, we defer this to future studies. 
\par
The rotational resistance can be computed in an analogous way, i.e., after computing the force distribution corresponding to a rotational motion of the filament, we compute the total torque, and its ratio to the angular velocity. 
\par
A consequence of the slenderness of rods is that the force density remains nearly uniform throughout the length of the rod in 3D flows for transnational motion, except near its two ends. 
%This uniform force distribution gives rise to the logarithmic correction to the mobility tensor \citep{Kim2013}. 
For rotational motion the force density varies linearly with arclength, changing sign at the center of mass, $s=0$. The end effects determine the next order corrections to the mobility/resistance, which depend on the detailed treatment of the rod geometry, e.g., assuming the rod is tapered or not \citep{johnson1980improved}. Thus, instead of solving a linear system of equations, one can obtain the leading order form of the mobility (or resistance) tensor, by assuming the force is uniform along the rod and simply evaluating the mobility as the ratio of the mean velocity to the total force. The equation for translational mobility is

\begin{equation}
\label{eq:FiberMobility3}
\boldsymbol{\chi} =L^{-2}\int_{-L/2}^{L/2}{d}s\int_{-L/2}^{L/2}
\mathbf{G} \left(\mathbf{X}(s)-\mathbf{X}(s^\prime)\right) {d}s^\prime.
\end{equation}

\begin{figure}
\begin{subfigure}[ ] 
{
\begin{minipage}{0.31\textwidth}
\centering
\hspace{0cm}
\vspace{0cm}
\includegraphics[width=0.8\textwidth]{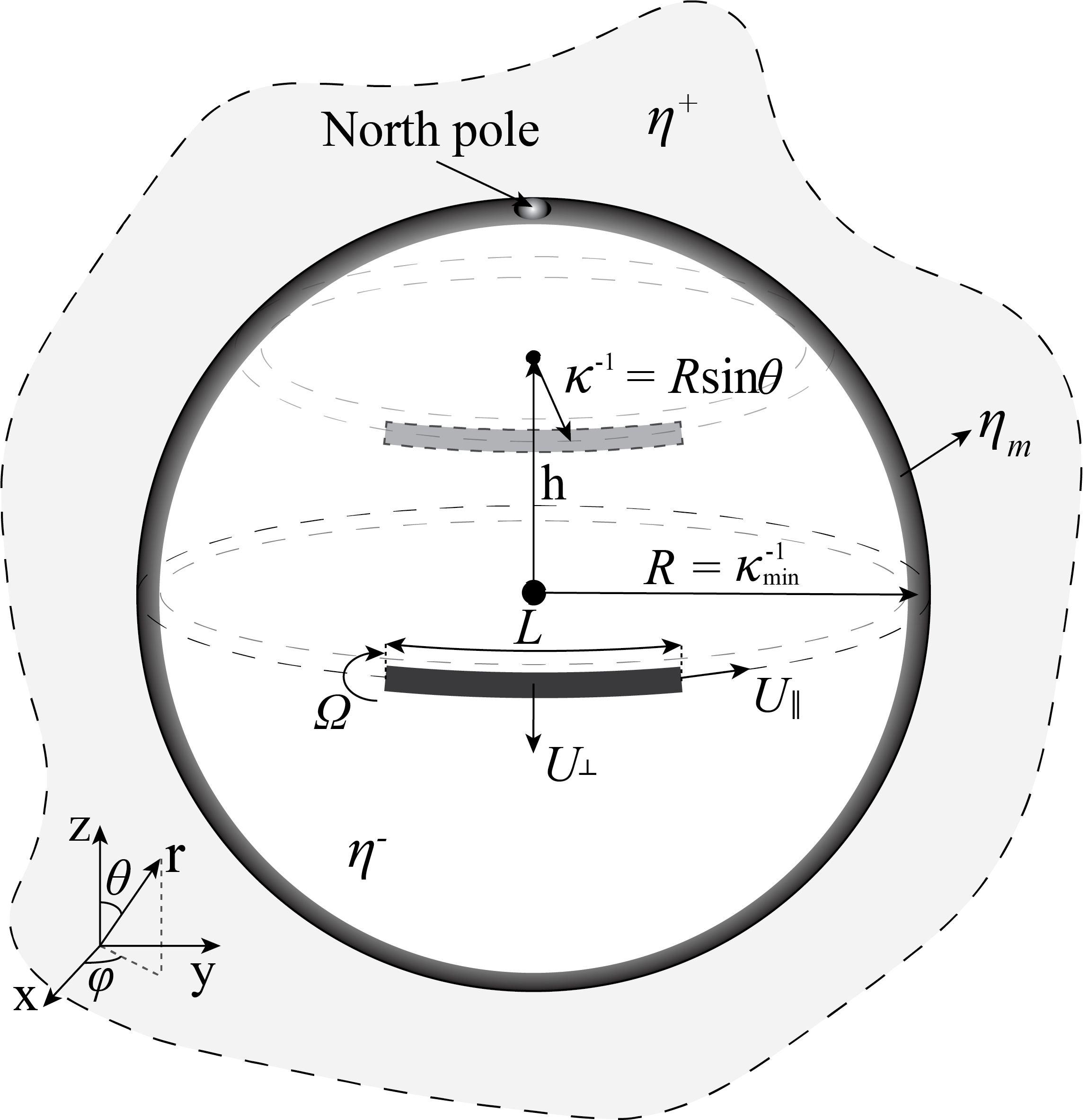}
\label{fig:illustration}
\end{minipage}
}
\end{subfigure}
\begin{subfigure}[] 
{
\begin{minipage}{0.30\textwidth}
\centering
\hspace{-0.5cm}
%\vspace{-1cm}
\includegraphics[width=0.8\textwidth]{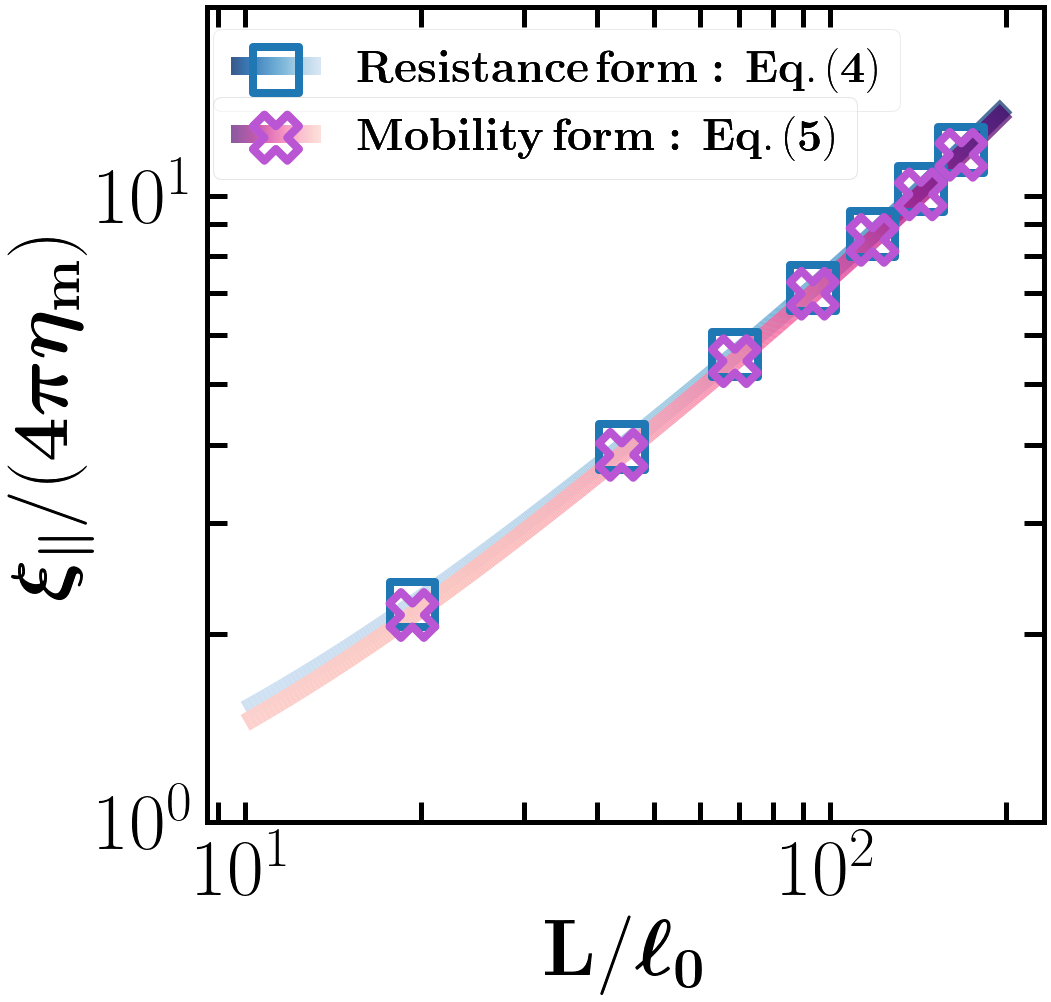}
\label{fig:two method_pa}
\end{minipage}
}
\end{subfigure}
\begin{subfigure}[] 
{
\begin{minipage}{0.30\textwidth}
\centering
\hspace{-0.5cm}
%\vspace{-1cm}
\includegraphics[width=0.8\textwidth]{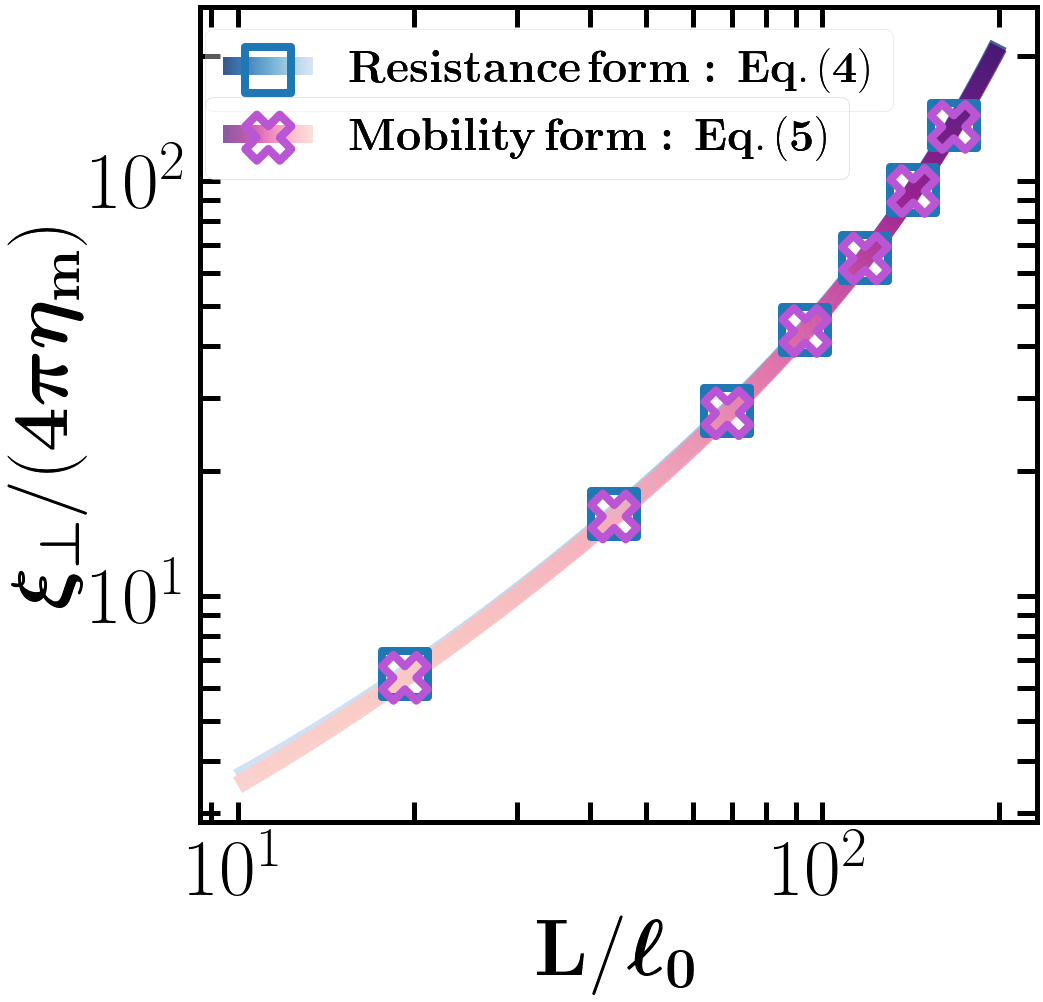}
\label{fig:two method_pe}
\end{minipage}
}
\end{subfigure}
\caption{(\textit{a}) An illustration of the filament motion on a spherical membrane of viscosity $\eta_m$ and surrounded on the interior and exterior sides by 3D fluid domains of shear viscosities $\eta^{-}$ and $\eta^{+}$. The filament dynamics is described by three modes of motion: translation along and perpendicular to its axis ($U_{\parallel}$ and $U_{\perp}$), and rotation ($\Omega$). The filament's curvature is at its minimum, when it is aligned along the equator ($\kappa_\text{min}=R^{-1}$). The results of this configuration are presented in \cref{sec:resist}. To model filaments of curvature $\kappa>\kappa_\text{min}$, they are placed at distance $h=\pm \sqrt{R^2-\kappa^{-2}}$ away from the equator. This configuration is studied in \cref{sec:off-equator}. 
(\textit{b,\,c}) The dimensionless resistance of a filament translating in parallel(\textit{b}) and perpendicular(\textit{c}) directions as a function of $L/\ell_{0}$, where $\ell_{0}/R=0.01$. The blue line (square dots) show the result of the numerical solutions to the resistance formulation given by \cref{eq:FiberMobility2}, while the red line (cross dots) show the numerical integration of the mobility formulation given by \cref{eq:FiberMobility3}, assuming a constant force density along the filament. The results are in excellent agreement with each other. 
\label{fig:illus_method}}
\end{figure}

Inverting the mobility tensor gives the resistance tensor: $\boldsymbol{\xi}=\boldsymbol{\chi}^{-1}$. \cref{fig:two method_pa,,fig:two method_pe} shows the predictions of the dimensionless resistance tensor in parallel and perpendicular directions as a function of $L/\ell_0$ when $\ell_{0}/R=0.01$ by solving \cref{eq:FiberMobility2} and \cref{eq:FiberMobility3}. The predictions from two methods are close with no more than $10\%$ and $20\%$ difference in parallel and perpendicular direction, respectively, for all the parameters space we have investigated. We provide a detailed analysis of the difference between two methods in Supplementary Materials. All the results reported hereafter were computed using \cref{eq:FiberMobility2}.
\par
Finally, we note that for all the results presented hereafter we set $\eta^{+}/\eta^{-}=1$. Varying this ratio only produced rather small $\mathcal{O}(1)$ changes in the final results and, thus, were not explored here to focus on large variations of the drag induced by flow confinements.

\section{Results} 
\subsection{Filament placed on the equator ( $\kappa=\kappa_\mathrm{min}=R^{-1}$)} \label{sec:resist}
\noindent
We begin by presenting the results of the filament at its minimum curvature $\kappa_\text{min}=1/R$. We note that applying a net force on filament bound to the spherical membrane produces a net torque on the membrane, which leads to a net rotational flow of the entire system \citep{Henle2010}. This effect is absent in a planar membrane \citep{samanta2021vortex}. The resistance and mobility functions are defined based on the \emph{relative} velocity of the filament with respect to the ambient fluid. As such, we subtract this rigid-body rotation from the velocity field of a moving filament in our computations of the resistance functions and surface flows. The same implementation was used by \citet{Henle2010}. The implementation details are provided in \cref{sec:appA}. 
\par
The filament's dynamics on a planar membrane is defined by two lengths: the filament's length, $L$, and SD length, $\ell_0$. Spherical geometry introduces another length: $R$. 
Constant filament curvature imposes an upper-bound on its length: $L<2\pi R$. We assume that the filament thickness, $a$, is significantly smaller than all other length-scales. The filament is modeled as a line with no thickness in the results presented here. In Supplementary Materials we show that the relative change in the computed resistance functions due to finite thickness of the filament at most scales with $\mathcal{O}(\epsilon)$, where $\epsilon=a/L$; hence, filament thickness has a negligible effect on the computed hydrodynamic functions. 

\begin{figure}
\begin{subfigure}[$\ell_0/R \ll 1$] 
{
\begin{minipage}{0.30\textwidth}
\centering
\hspace{-0.2cm}
\includegraphics[width=0.9\textwidth]{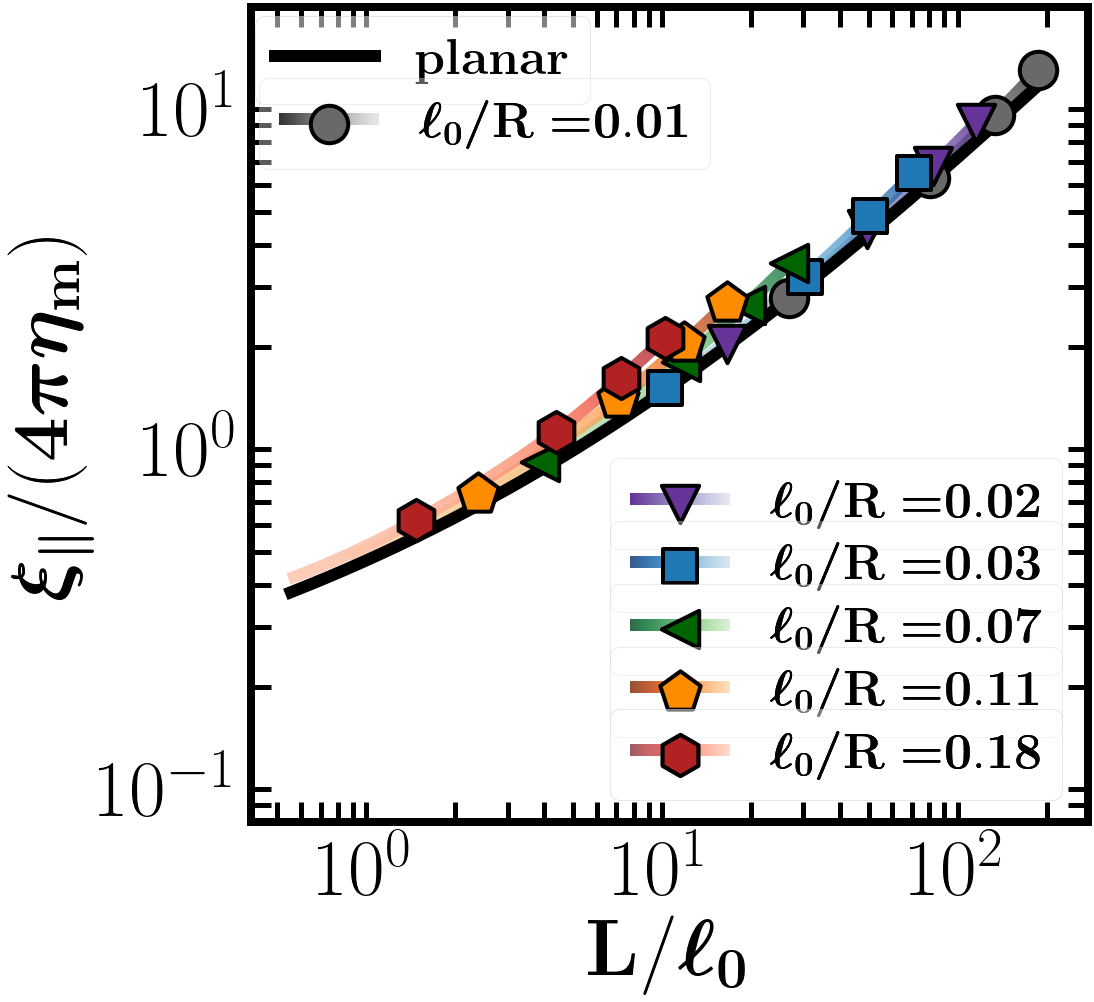}
\label{fig:pa_001}
\end{minipage}
}
\end{subfigure}
\begin{subfigure}[$\ell_0/R \gg 1$] 
{
\begin{minipage}{0.30\textwidth}
\centering
\hspace{-0.85cm}
\includegraphics[width=0.9\textwidth]{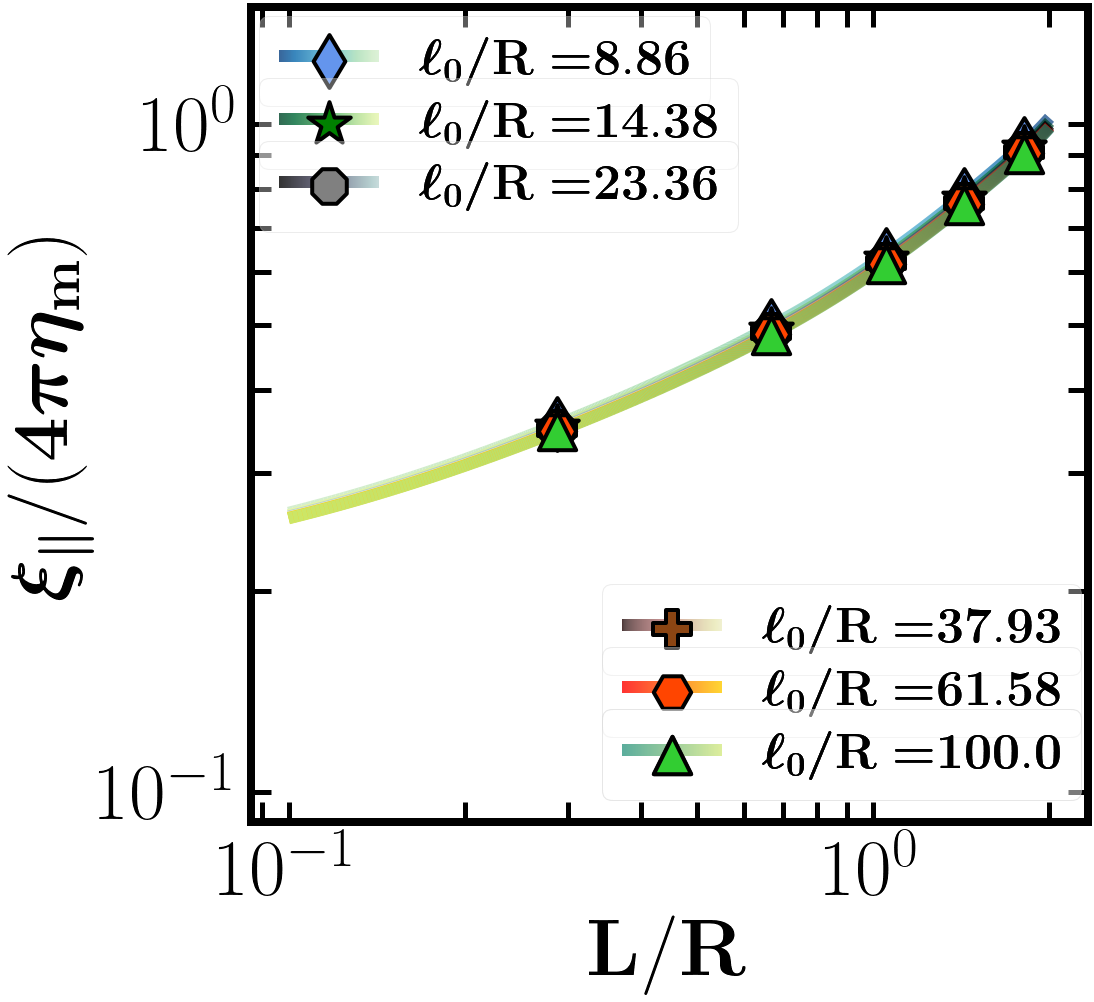}
\label{fig:pa_100}
\end{minipage}
}
\end{subfigure}
\begin{subfigure}[$10^{-2} \le \ell_0/R \le 10^{2}$] 
{
\begin{minipage}{0.30\textwidth}
\centering
\hspace{-1.3cm}
\includegraphics[width=0.9\textwidth]{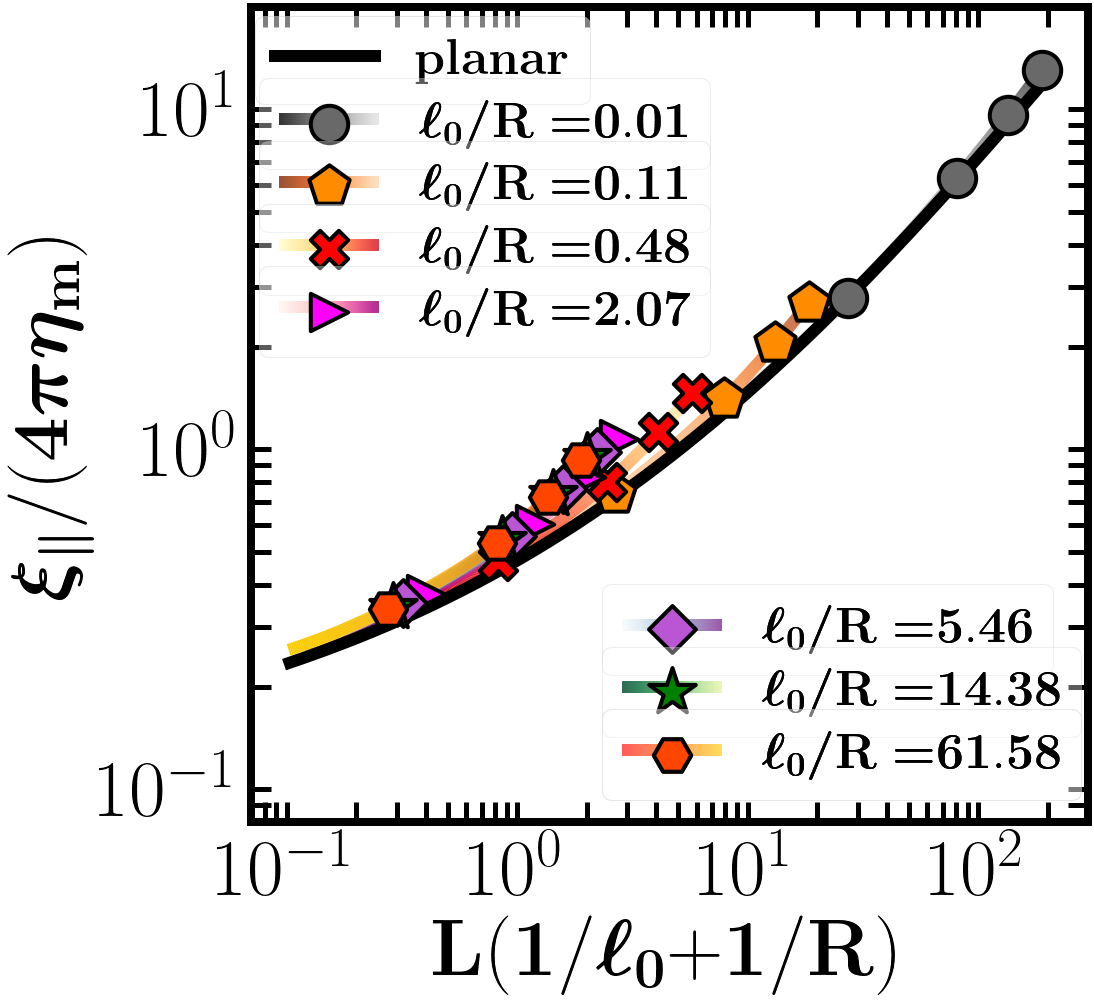}
\label{fig:pa_1}
\end{minipage}
}
\end{subfigure}
\begin{subfigure}[$\ell_0/R \ll 1$] 
{
\begin{minipage}{0.30\textwidth}
\centering
\hspace{-0.2cm}
\includegraphics[width=0.9\textwidth]{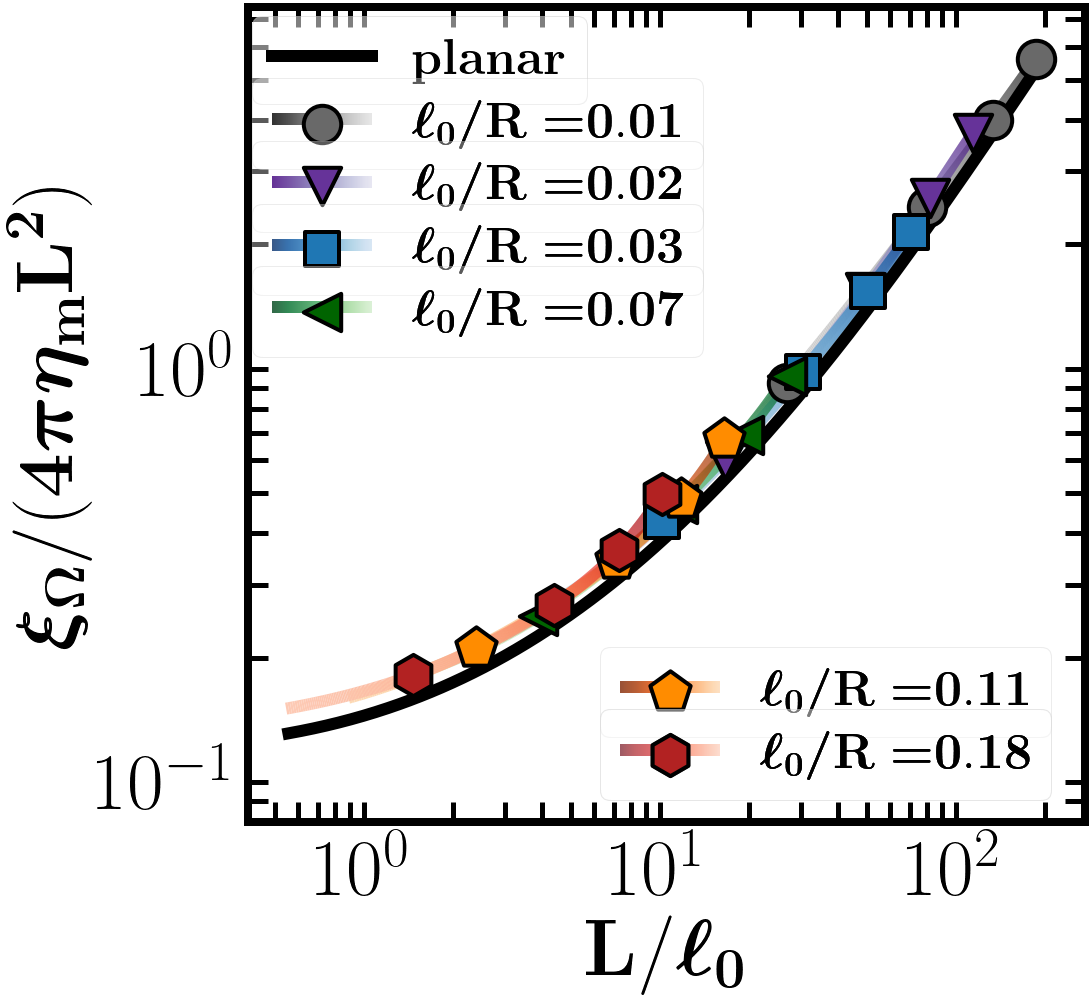}
\label{fig:ro_001}
\end{minipage}
}
\end{subfigure}
\begin{subfigure}[$\ell_0/R \gg 1$] 
{
\begin{minipage}{0.30\textwidth}
\centering
\hspace{-0.85cm}
\includegraphics[width=0.9\textwidth]{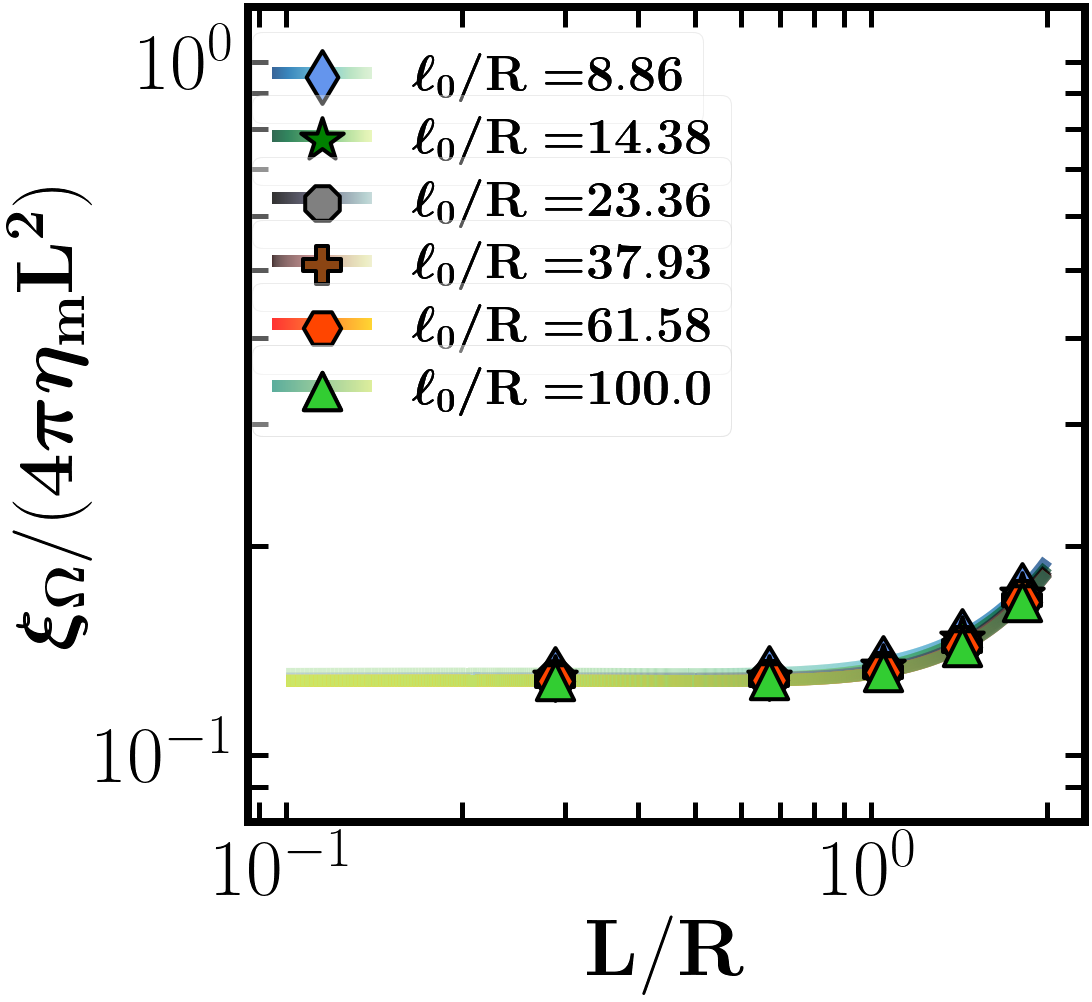}
\label{fig:ro_100}
\end{minipage}
}
\end{subfigure}
\begin{subfigure}[ $10^{-2} \le \ell_0/R \le 10^{2}$] 
{
\begin{minipage}{0.30\textwidth}
\centering
\hspace{-1.3cm}
\includegraphics[width=0.9\textwidth]{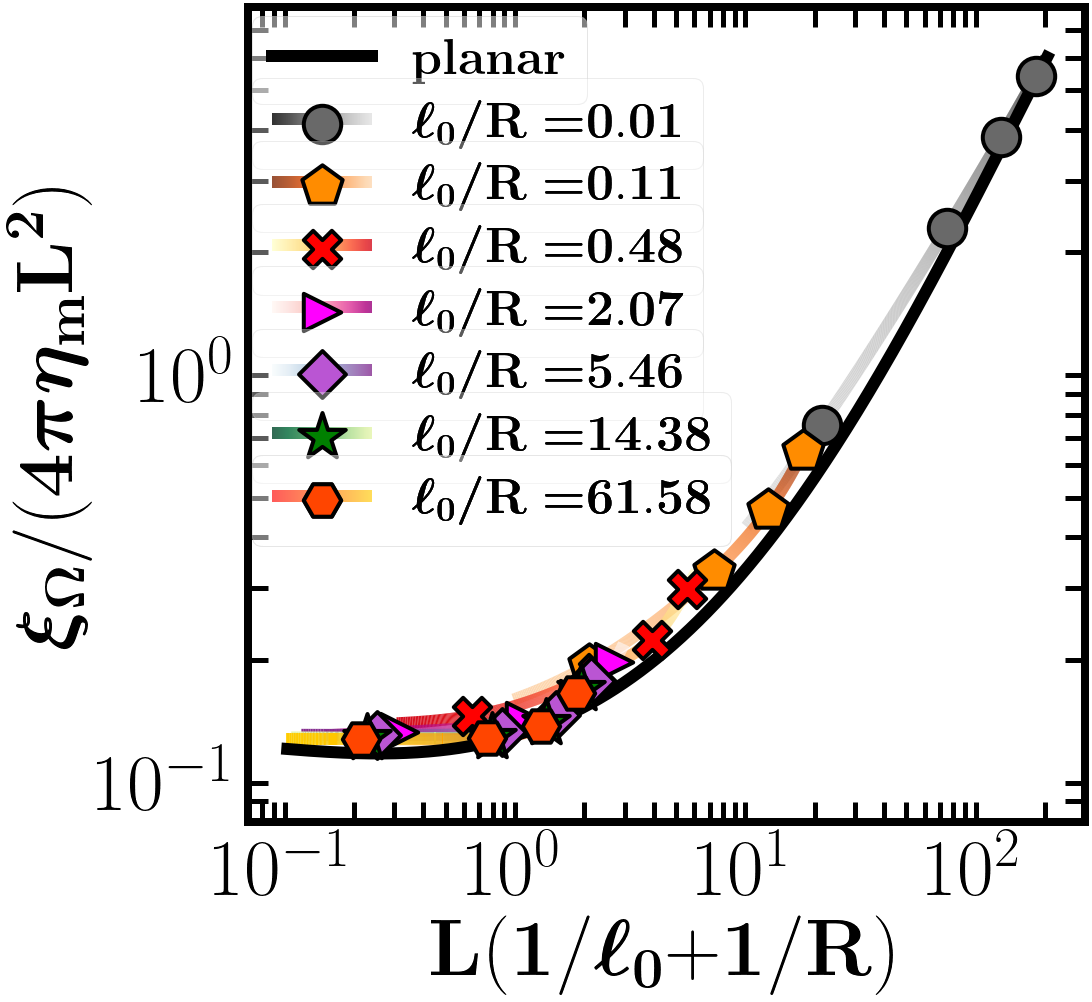}
\label{fig:ro_1}
\end{minipage}
}
\end{subfigure}
\begin{subfigure}[$\ell_0/R \ll 1$] 
{
\begin{minipage}{0.30\textwidth}
\centering
\hspace{-0.1cm}
\includegraphics[width=0.9\textwidth]{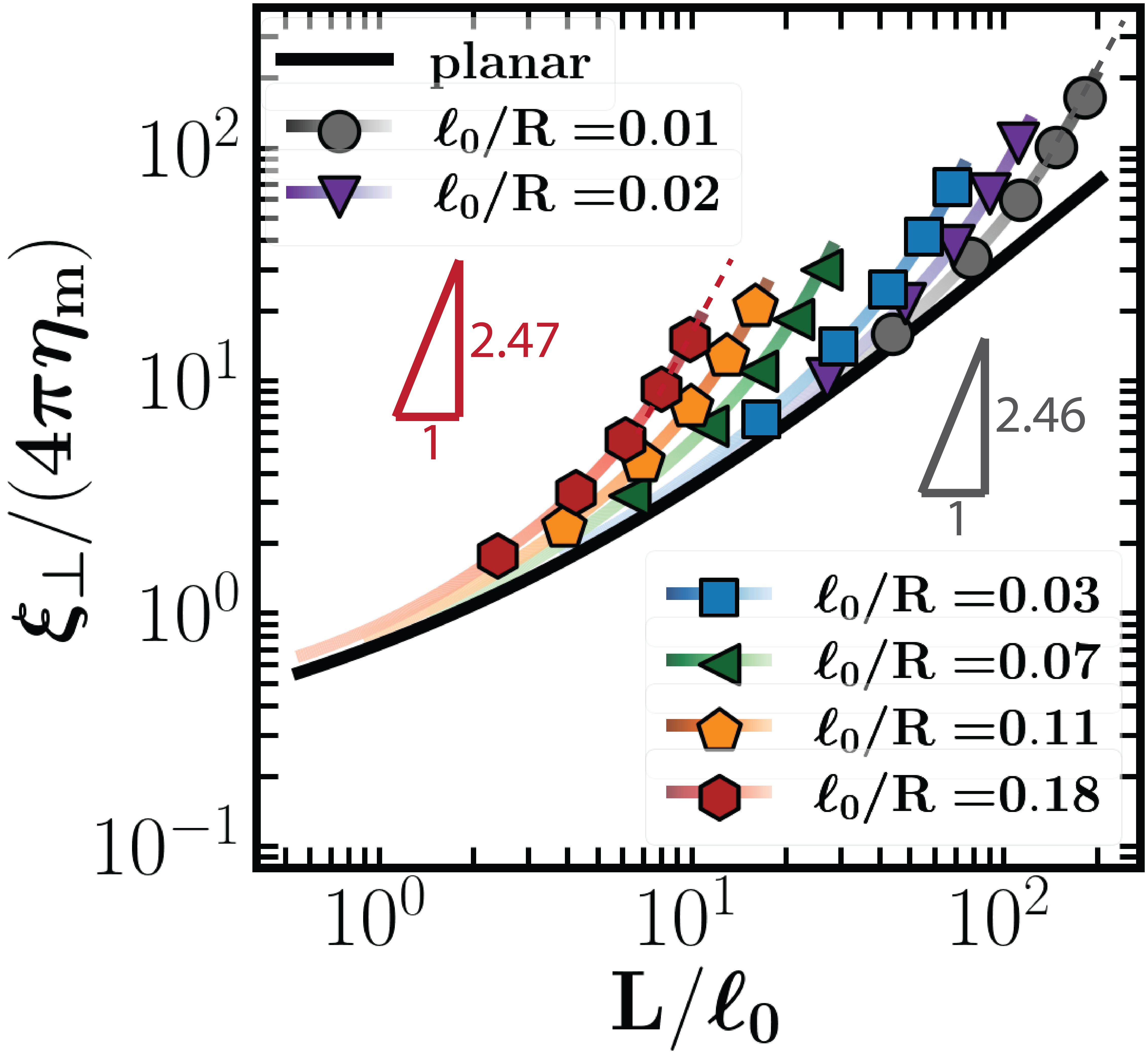}
\label{fig:pe_001}
\end{minipage}
}
\end{subfigure}
\begin{subfigure}[$\ell_0/R \gg 1$] 
{
\begin{minipage}{0.30\textwidth}
\centering
\hspace{-0.75cm}
\includegraphics[width=0.9\textwidth]{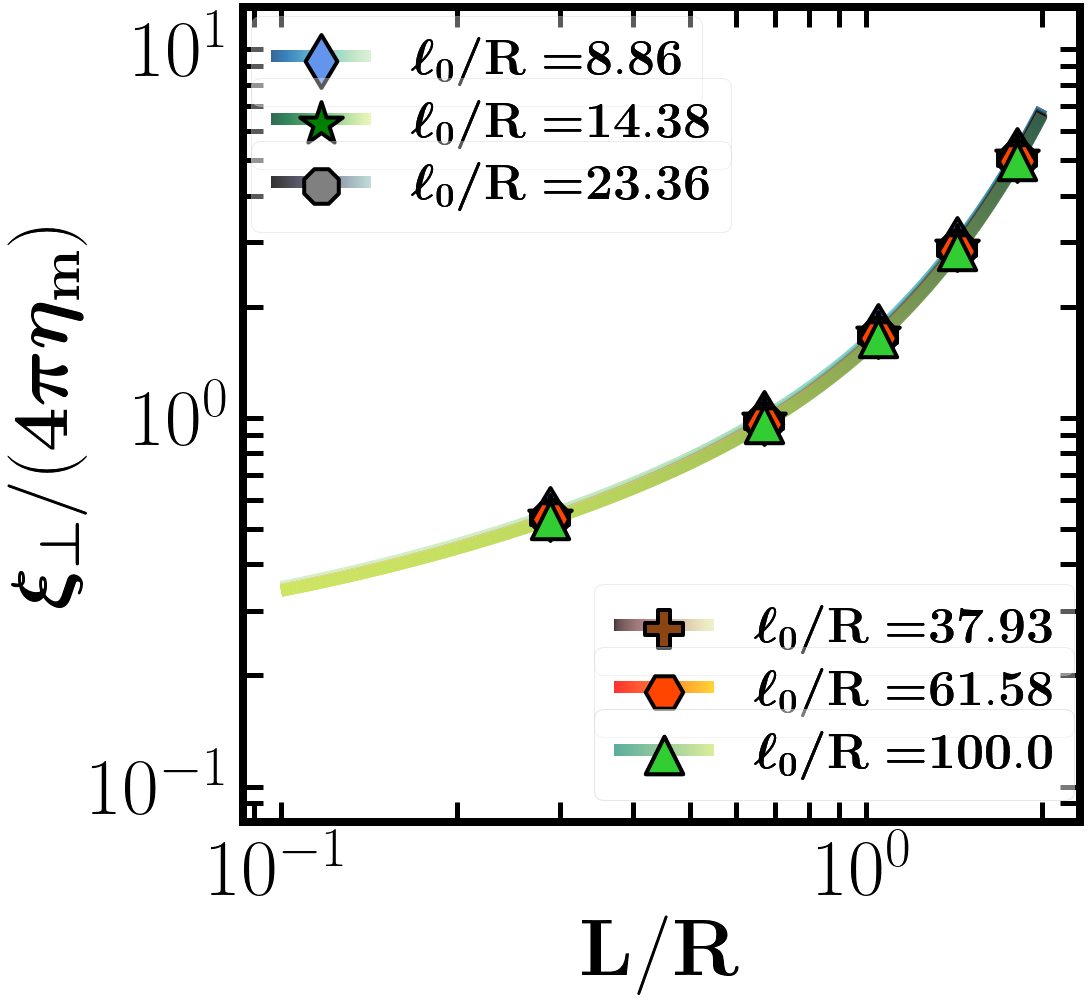}
\label{fig:pe_100}
\end{minipage}
}
\end{subfigure}
\begin{subfigure}[ $10^{-2} \le \ell_0/R \le 10^{2}$] 
{
\begin{minipage}{0.30\textwidth}
\centering
\hspace{-1.2cm}
\includegraphics[width=0.9\textwidth]{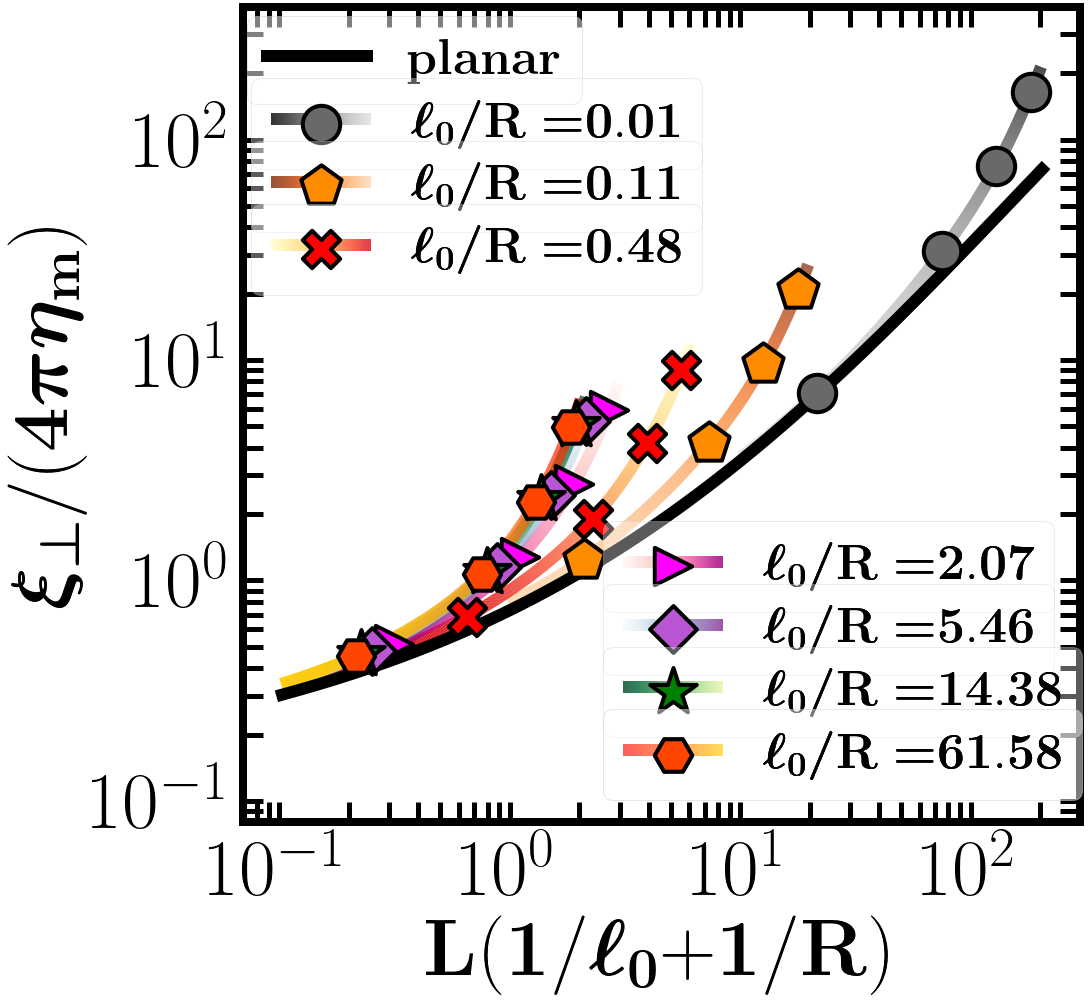}
\label{fig:pe_1}
\end{minipage}
}
\end{subfigure}
\caption{The parallel(upper row), perpendicular(lower row) and rotational(middle row) resistance as a function of $L/\ell_{0}$ and $L/R$ in different limits of $\ell_0/R$. The left column and middle column correspond to $\ell_{0}\ll{R}$ and $\ell_{0}\gg{R}$, respectively. The right column includes all data in the range $10^{-2}\le \ell_{0}/R\le 10^2$. The solid black lines in (\textit{a}), (\textit{c}), (\textit{d}), (\textit{f}), (\textit{g}), and (\textit{i}) represent the associated resistance values in planar membranes \citep{Levine2004}. The variations of $L/R$ for each choice of $\ell_0/R$ is visualized by changing the color from light (left-end of each plot) to dark (right-end of each plot) with  increasing $L/R$. 
\label{fig:resist}}
\end{figure}

We study translational resistance, in parallel and perpendicular directions, and rotational resistance in three regimes: $\ell_0/R\ll 1$, $\ell_0/R\sim \mathcal{O}(1)$, and $\ell_0/R\gg 1$. The results are presented in \cref{fig:resist}. In all cases the variations of the resistance are
presented as the length of the filament is changed, while fixing $\ell_0/R$. Thus, $L/R$ varies in each individual line in \cref{fig:resist}. 
These variations are 
visualized by varying the transparency of each color in the plot from more transparent (lighter) in smaller $L/R$ to less transparent (darker) in larger $L/R$. 
\par
As shown in \cref{fig:pa_001}, in the limit of $\ell_0/R\ll 1$ the resistance plots in parallel direction for different values of $\ell_0/R$ nearly collapse to the resistance values of a planar membrane at $L/R\ll 1$, and show relatively weak positive deviations which is at most $40\%$ from this curve. 
The positive deviations result from the flow confinement effects in a closed spherical geometry and, thus, increase with $L/R$. The details are provided in Supplementary Materials. 
%This effect is absent in 2D free space planar membranes and cannot be captured in the simple rescaling of $\ell^\star$. 
\par
In the opposite limit of $\ell_0/R\gg 1$, shown in \cref{fig:pa_100}, the resistance (and mobility) become independent of $\ell_0$, resulting in nearly perfect collapse of all the results when plotted against $L/R$. Both limits can be understood by noting that the length-scale that determines the flows and the hydrodynamic functions is the length-scale over which the momentum is transported from the membrane to the surrounding 3D fluid domains, which we refer to as $\ell^\star$ hereafter. In case of $\ell_0/R \ll 1$, the momentum transport occurs in $\ell^\star=\ell_0$, and the results closely follow those of a planar membrane. In the other limit of $\ell_0/R\gg 1$, the momentum transfer occurs over the dimension of the spherical geometry, $R$, which is smaller than $\ell_0$. As such, the results are independent of $\ell_0$. 
\par
Based on this understanding, we combine the results of \cref{fig:pa_001,,fig:pa_100} into \cref{fig:pa_1}, by rescaling the length in $x-$axis with $\ell^\star=\left(\ell_0^{-1}+R^{-1}\right)^{-1}$, which asymptotes to the expected limits. With this rescaling, the computed parallel resistance for a wide range of $\ell_0/R$ collapse to its planar membrane values when $L/R<1$ (lighter colors), and show weak positive deviations at larger $L/R$ values.
\par
\cref{fig:ro_001,,fig:ro_100,,fig:ro_1} present the rotational resistance at the same limits as the parallel motion. As it can be seen, similar to the parallel resistance, the rotational resistance closely matched that of a planar membrane, when the filament's length is appropriately made dimensionless by $\ell^\star$. 

Finally, in \cref{fig:pe_001,,fig:pe_100,,fig:pe_1} we present the variations of translational drag in perpendicular direction as a function of the filament's length in the same regimes. When $\ell_{0}/R\ll1$ and $L/R\ll1$, the computed drag coefficients converge to the planar membrane values. In contrast to the parallel and rotational drag, we observe strong positive deviations from the planar membrane results with increasing $L/R$. This demonstrates that flow confinement have a significantly stronger effect on $\xi_\perp$ than $\xi_\parallel$ and $\xi_\Omega$. In the limit of $\ell_0/R \gg 1$, similar to the previous cases, the dynamics becomes independent of $\ell_0$, resulting in the collapse of all data when plotted against $L/R$. However, these variations are significantly stronger than those in parallel and rotational resistance functions.
As expected rescaling the length with $\ell^\star$ cannot collapse the data.
Note that to move in perpendicular direction and to remain bound to the membrane, the filament needs to deform and change its curvature. These deformations are induced by membrane-filament interactions in the radial direction, and produce no net force in the surface of the sphere. Thus, the computed instantaneous perpendicular resistance remains unchanged by these deformations. 
% as $\xi_\perp$ increases superlinearly with $L/R$ when $L/R >1$.  
\par
\begin{figure}
\begin{subfigure}[] 
{
\begin{minipage}{0.3\textwidth}
\centering
\hspace{0.0cm}
\includegraphics[width=0.8\textwidth]{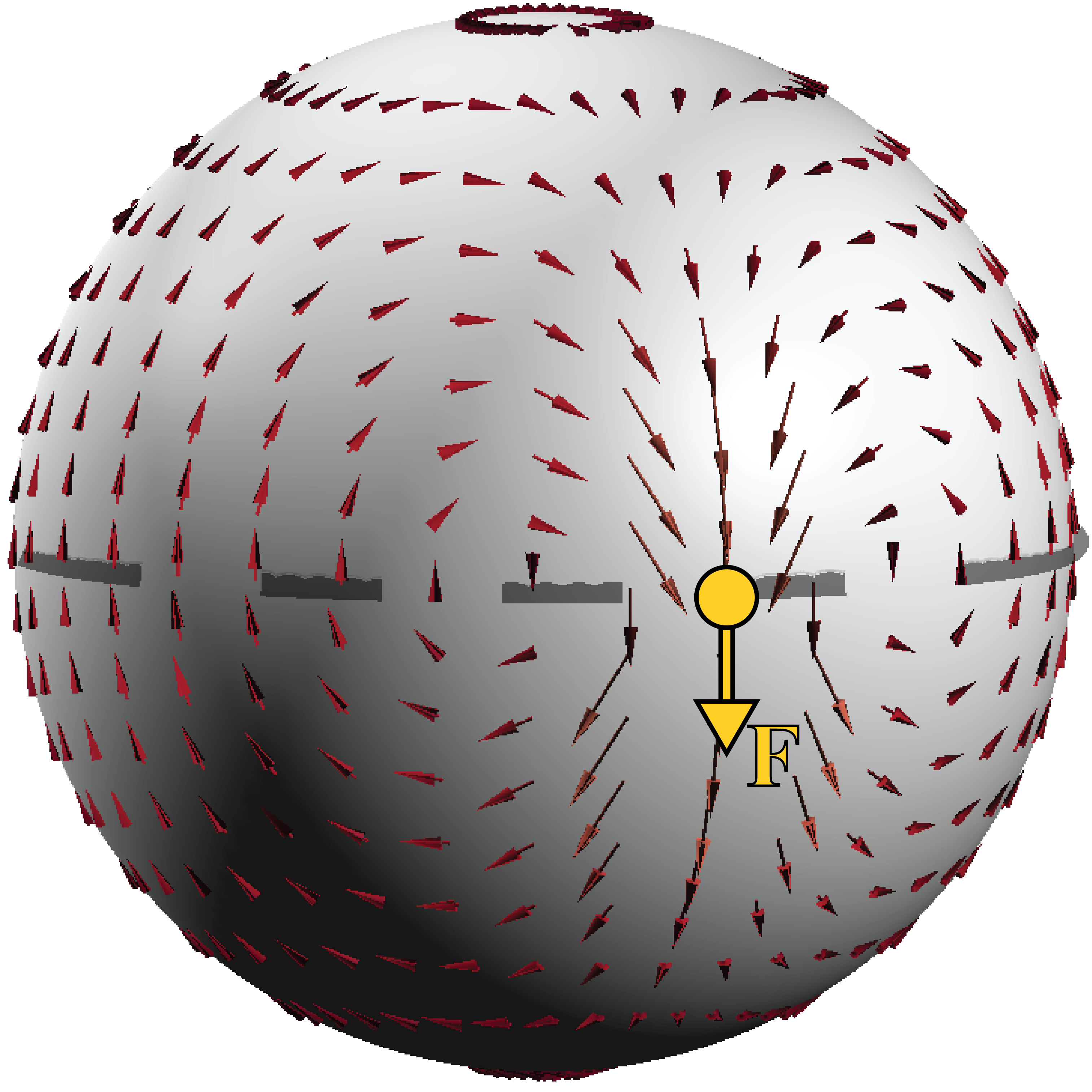}
\label{fig:flow_pe_1_point}
\end{minipage}
}
\end{subfigure}
\begin{subfigure}[] 
{
\begin{minipage}{0.3\textwidth}
\centering
\hspace{0.0cm}
\vspace{-0.0cm}
\includegraphics[width=0.8\textwidth]{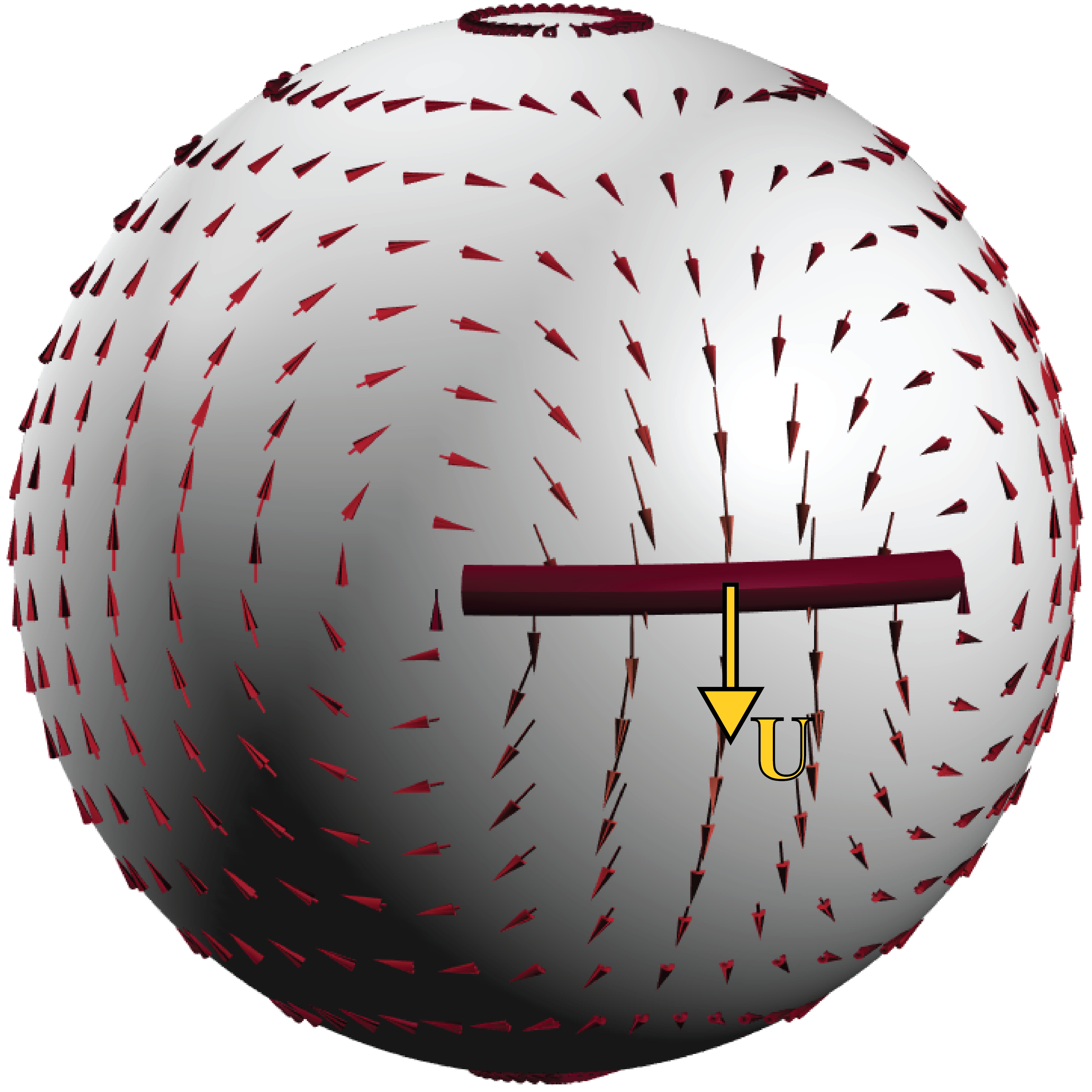}
\label{fig:flow_pe_1_irr}
\end{minipage}
}
\end{subfigure}
\begin{subfigure}[] 
{
\begin{minipage}{0.3\textwidth}
\centering
\hspace{0.0cm}
\vspace{-0.0cm}
\includegraphics[width=0.8\textwidth]{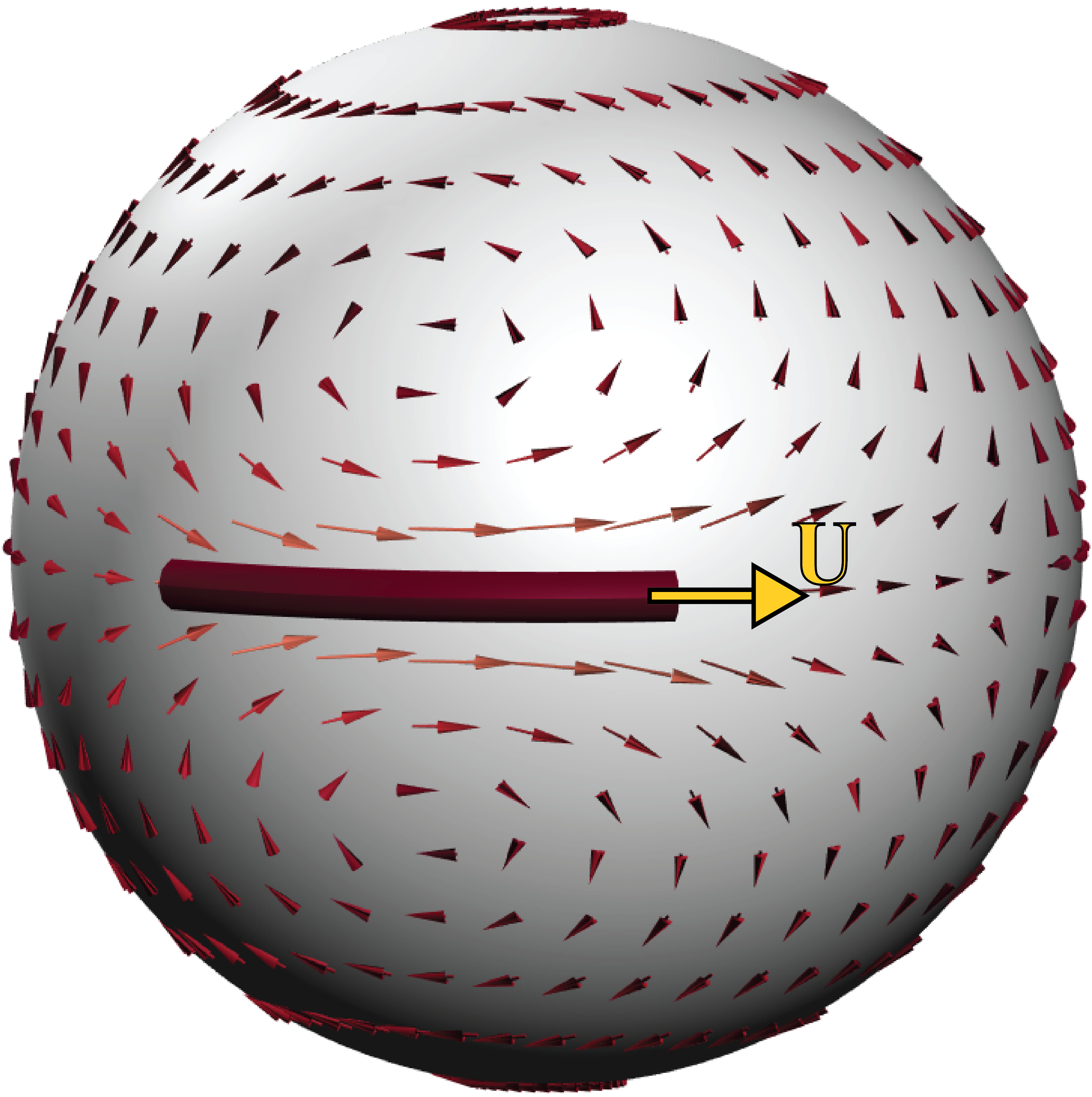}
\label{fig:flow_pa_1_irr}
\end{minipage}
}
\end{subfigure}
\caption{(\textit{a}) The flow field induced by a point-force in the perpendicular direction at ($\theta=\pi/2$, $\phi=0$) for the choice of $\ell_{0}/R=1$. (\textit{b}) The flow field induced by a filament moving with a constant velocity in the perpendicular direction for the choice of $\ell_{0}/R=1$ and $L/R=1$. (\textit{c}) The flow field induced by a filament moving with a constant velocity in the parallel direction for the choice of $\ell_{0}/R=1$ and $L/R=1$.
\label{fig:flow}}
\end{figure}

\cref{fig:flow_pe_1_point} shows the flow generated by a point-force placed on the equator and pointing towards the south-pole. Note that we have subtracted the flows due to the pure rotational motion of the membrane and surrounding fluids, as discussed earlier in this section. As it can be seen, the boundedness of spherical geometry produces strong flow reversal and re-circulation regions as we move in the direction perpendicular to the point-force ($\hat{\mathbf{\phi}}$), whereas the flows remain parallel to the point-force, when moving in the direction parallel to the force ($\hat{\mathbf{\theta}}$).  
\par
\cref{fig:flow_pe_1_irr} shows the velocity fields induced by filament's motion in the perpendicular direction. 
The flows are qualitatively similar to the flow induced by a point-force with the difference that the flow reversal regions are now pushed to areas near the filament's ends. 
As we discussed in section \cref{sec:num_method}, a filament moving with a constant velocity can be modeled as 
a collection of smaller segments all moving with the same velocity, where the coupling between the segments is through their HIs. The flow generated by a segment is similar to the flow due to a point-force, shown in \cref{fig:flow_pe_1_point}. Thus, when the filament is moving in perpendicular direction, the motion of one segment can generate flows pointing in the opposite direction of motion on other segments, i.e., HIs cause anti-correlated motions. As a result, more force is needed to maintain the same velocity across all segments, leading to an increase in the total drag compared to the free draining limit. 
\par
\cref{fig:flow_pa_1_irr} shows the velocity field induced by the filament's parallel motion.
In this case, the motion of a segment produces flows in the same direction as the motion on other filament's segments, i.e.,
HIs cause correlated motion, leading to a reduction in the total drag, compared to the free draining limit. 
\par
\begin{figure}
\begin{subfigure}[ ] 
{
\begin{minipage}{0.48\textwidth}
\centering
\hspace{-0.6cm}
\vspace{-0.0cm}
\includegraphics[width=0.9\textwidth]{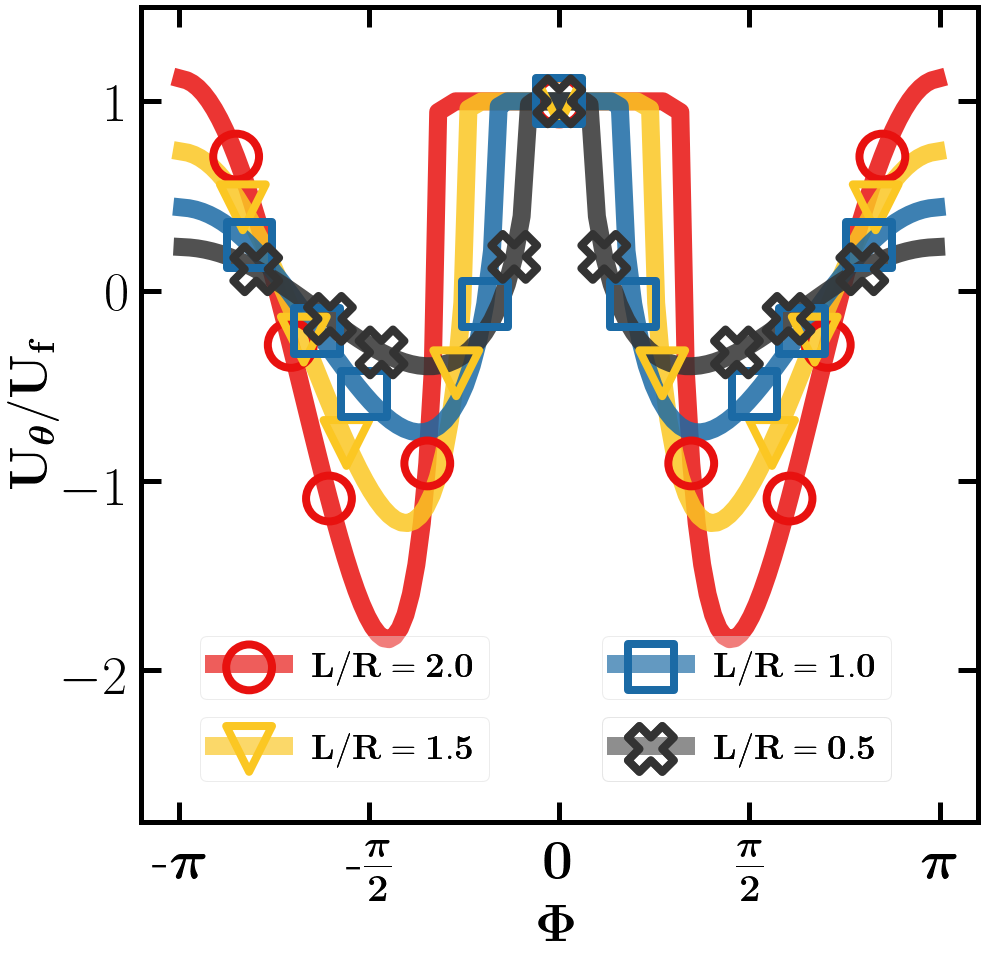}
\label{fig:equator}
\end{minipage}
}
\end{subfigure}
\begin{subfigure}[ ] 
{
\begin{minipage}{0.48\textwidth}
\centering
\hspace{-0.6cm}
\includegraphics[width=0.9\textwidth]{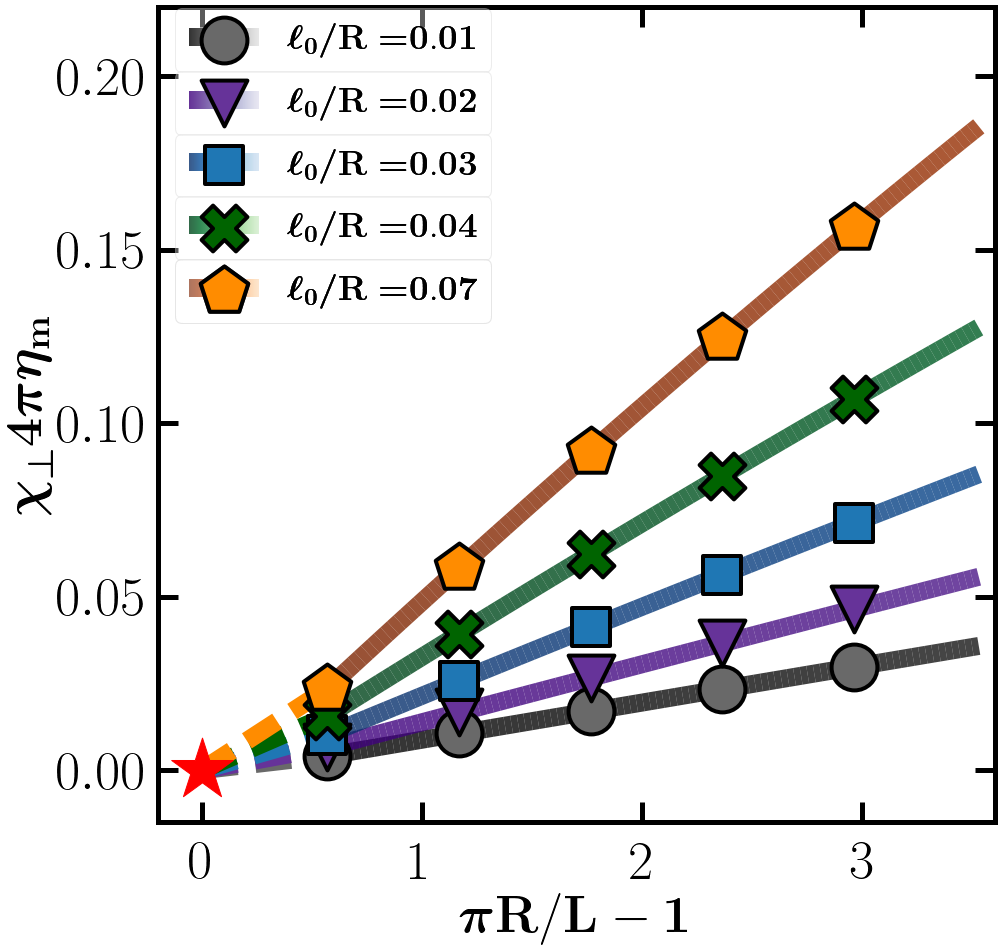}
\label{fig:peM_001}
\end{minipage}
\label{fig:MobilityGap}}
\end{subfigure}
\caption{(\textit{a}) The polar velocity distributed in the azimuthal direction along the equator, normalized by the filament velocity for the choice of $\ell_{0}/R=1$. The different colors/symbols represent different ratios of $L/R$. (\textit{b}): The mobility of filament in perpendicular direction as a function of gap distance, $\pi R/L-1$, for different values of $\ell_0/R$. 
%The different colors/symbols represent different ratios of $\ell_{0}/R$. 
\label{fig:peM}}
\end{figure}

In \cref{fig:equator} we quantify these variations of surface velocity in perpendicular direction along the equator, and at different filament lengths. The results are obtained for the choice of $\ell_0/R=1$; again, we have removed the pure rotation component of the velocity field. The flow gradients are the strongest near the filament's ends, and they strongly increase with the filament's length. These flow gradients result in large traction forces near the filament's ends and a strong increase in the total drag. % and appear to diverge as $L/R \to \pi$. 
\par
\cref{fig:peM_001} shows the perpendicular mobility $\chi_\perp=\xi_\perp^{-1}$ as a function of the gap size, for different values of $\ell_0/R$, where we define the gap size as the distance between membrane's half -perimeter and filament's length: $\pi{R}/L-1$, which non-dimensionalized by filament length. In all plots the mobility approaches zero, $\chi_\perp\to 0$, corresponding to $\xi_\perp \to \infty$, as $L \to \pi{R}$. These results are in line with the results of \cref{fig:pe_001} and \cref{fig:equator}, demonstrating the rapid growth of $\xi_\perp$ with filament's length when $L/R>1$. Obtaining the asymptotic form of mobility requires a more in-depth analysis and careful numerical evaluations, which we do not pursue in this study. 

\subsection{Filament placed away from the equator ($\kappa>R^{-1}$)} \label{sec:off-equator}
%\noindent
%\textbf{\textcolor{red}{behavior of the green's function.}}
%\par
%\noindent
Thus far we have focused on the case where the filament's curvature is at its minimum, $\kappa_\text{min}=1/R$.  
Next, we ask how does changing filament's curvature, independently of the membrane geometry, affect its dynamics? This introduces another dimensionless parameter $\kappa R$, which is the ratio of filament's curvature to its minimum allowed curvature on that membrane. 
Due to the mismatch between the membrane and filament curvatures, 
the filament cannot undergo pure rotational motion, while remaining bound to the membrane. 
The same conditions apply to motion along perpendicular direction, which makes defining and computing $\xi_R$ and $\xi_\perp$ problematic when $\kappa \neq 1/R$. Because of this complexity, we only focus on the well-defined dynamics in the parallel direction.

\begin{figure}
\begin{subfigure}[ ] 
{
\begin{minipage}{0.26\textwidth}
\centering
\hspace{0.1cm}
\includegraphics[width=0.9\textwidth]{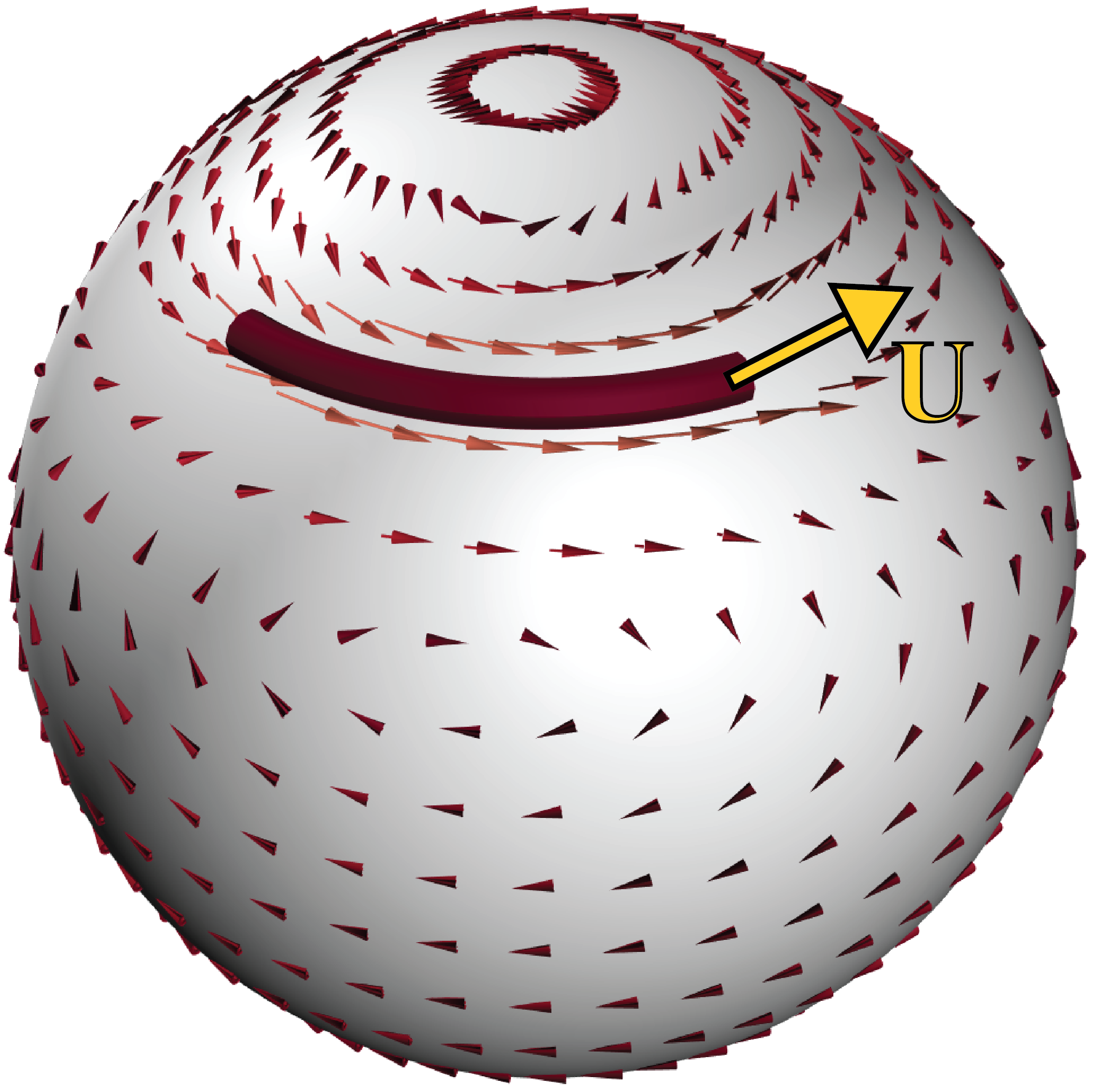}
\label{fig:flow_pa_1_off}
\end{minipage}
}
\end{subfigure}
\begin{subfigure}[ ] 
{
\begin{minipage}{0.32\textwidth}
\centering
\hspace{0cm}
\includegraphics[width=0.9\textwidth]{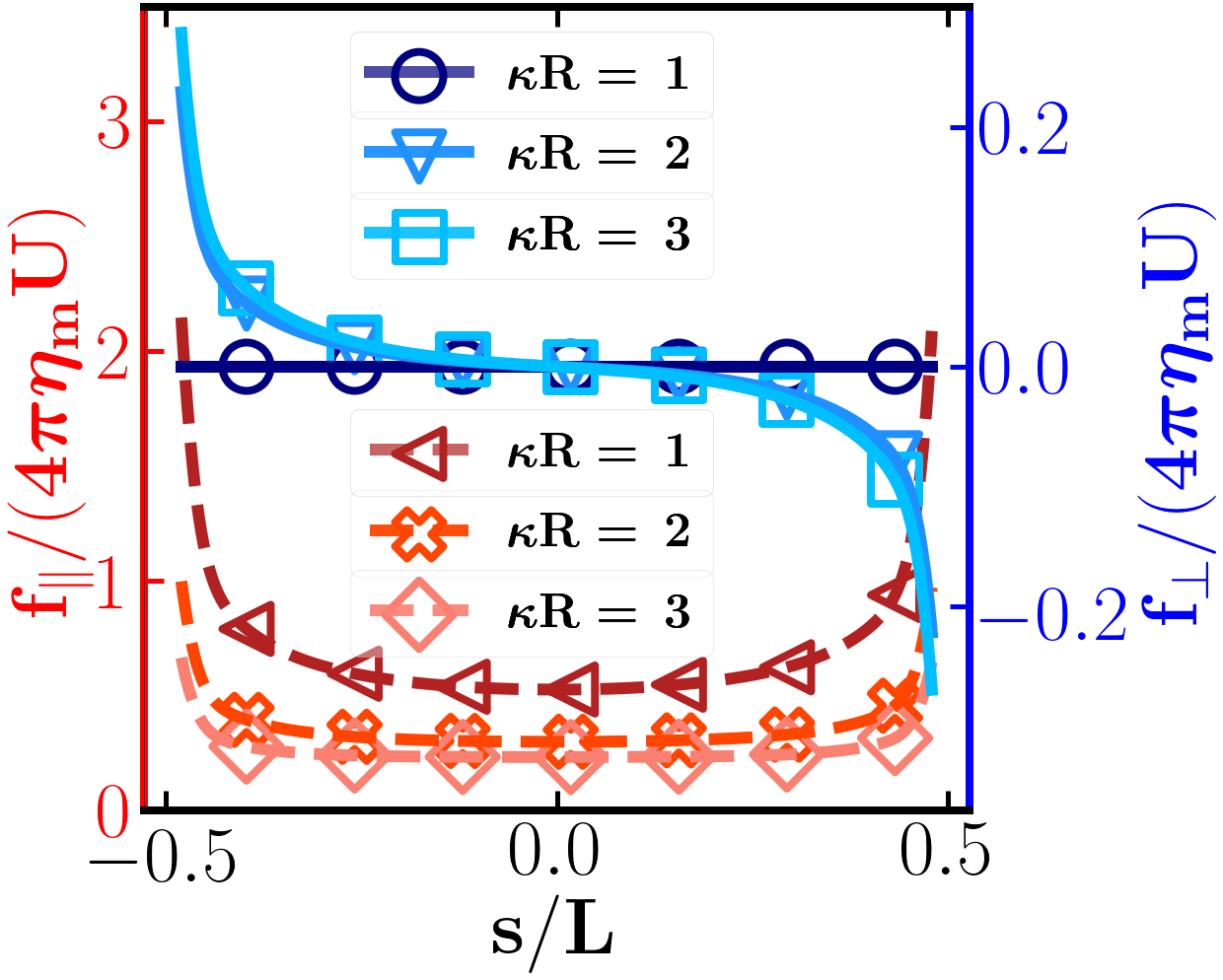}
\label{fig:off_f_1}
\end{minipage}
}
\end{subfigure}
\begin{subfigure}[ ] 
{
\begin{minipage}{0.29\textwidth}
\centering
\hspace{-0.5cm}
\includegraphics[width=0.9\textwidth]{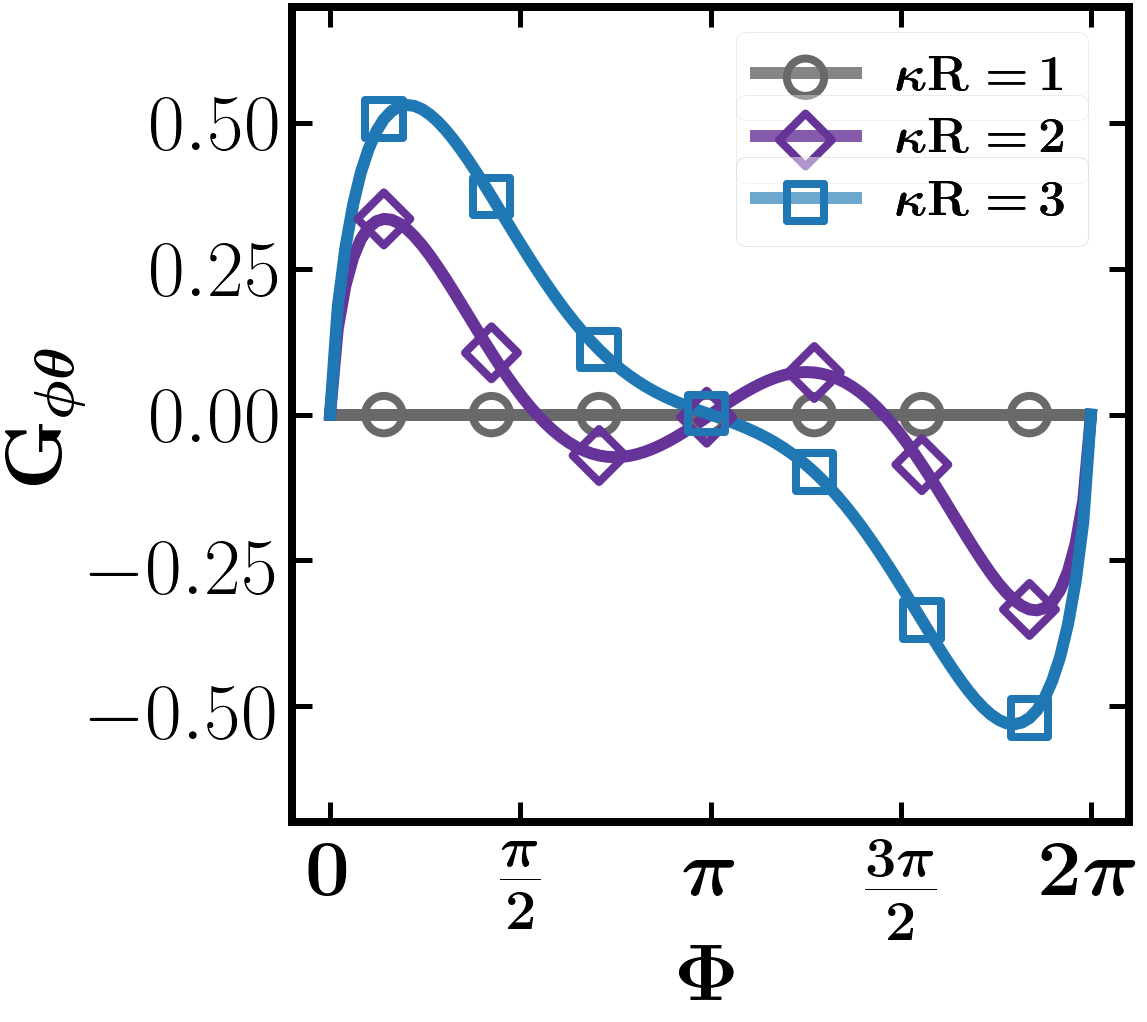}
\label{fig:green_offd_1}
\end{minipage}
}
\end{subfigure}
\caption{(\textit{a}) The flow field induced by the parallel motion of a filament placed at $h=R/\sqrt{2}$ away from the equator ($\kappa=\sqrt{2}R^{-1}$), when $\ell_{0}/R=1$ and $L/R=1$. (\textit{b}) The distribution of force per unit length along the filament in parallel and perpendicular directions ($\mathbf{f}_{\parallel,\perp}$) at different latitudes (curvatures), when $\ell_{0}/R=1$ and $L/R=1$. (\textit{c}) Variations of $G_{\phi\theta}$ vs $\phi$, at different latitudes/curvatures ($\theta_{0}$), when the source point is located at $(\theta_0,\phi_0=0)$ and the target is at $(\theta_0,\phi)$. The results are presented for $\ell_{0}/R=1$; see \cref{sec:appA} for the full expression of the Green's function. 
\label{fig:Green}}
\end{figure}

\cref{fig:flow_pa_1_off} shows the surface fluid flows induced by the parallel motion of a filament placed in-between the equator and the north pole, when $\ell_0/R=1$ and $L/R=1$ and $\kappa R=\sqrt{2}$. As expected, the flows are asymmetric around the filament. This flow asymmetry produces nonzero distribution of forces along the filament in perpendicular directions, $f_\perp$, which are shown in \cref{fig:off_f_1} for different values of $\kappa R$. 
Note that the parallel force distribution remains  uniform (except near the ends) and symmetric around the filament's mid-point ($s=0$), while the  
perpendicular force is anti-symmetric around  $s=0$, and increases in magnitude as filament nears the pole. The total torque normal to the spherical boundary scales with $\int_{-L/2}^{L/2} sf_{\perp}\mathrm{d}s$. The integrand is an even function, resulting in a nonzero torque on the filament \footnote{This torque changes sign as the filament is placed on the southern hemisphere.}. In other words, placing the filament away from the equator (increasing the filament's curvature) results in the coupling of its rotational and parallel motions. 
\par
It is also useful to describe this coupling of motions by inspecting the
mathematical structure of the fundamental solutions. \cref{fig:green_offd_1} shows variations of off-diagonal component of periodic Green's function, $G_{\theta\phi}$ along the azimuthal direction at different distances from the equator. As it can be seen, $G_{\theta\phi}$ is anti-symmetric and increases in magnitude as we move towards the poles. Since $G_{\phi\theta}$ is an odd function with respect to $\phi$, applying a uniform force in parallel ($\hat{\mathbf{\phi}}$) direction of the filament in \cref{eq:sbt1} produces a rotational flow around it; see full expressions in \cref{sec:appA}.
\par
%As shown in \cref{fig:off_F_1}, the force distribution along the filament as filament moving in parallel direction but at different latitude. The parallel force is insensitive to the latitude which suggest that the parallel resistance is insensitive to latitude coherent to we stated before that parallel motion is not restricted to geometrical confinement. While the perpendicular force is identically zero on the equator but anti-symmetrical off-equator which is consistent to the behavior of the off-diagonal component of the Green's function.  

\begin{figure}
\begin{subfigure}[$\ell_0/R=0.01$] 
{
\begin{minipage}{0.30\textwidth}
\centering
%\hspace{-0.2cm}
\includegraphics[width=0.9\textwidth]{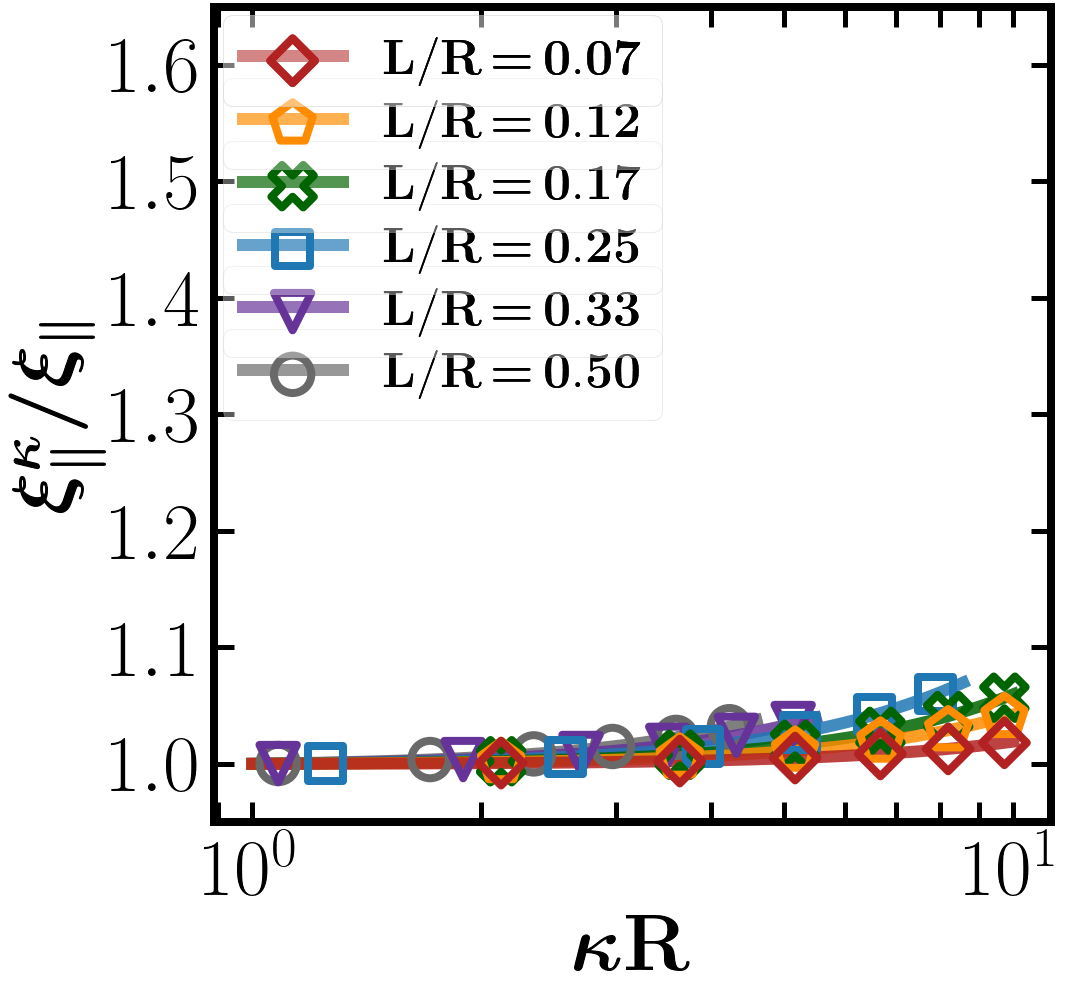}
\label{fig:pa_off_001}
\end{minipage}
}
\end{subfigure}
\begin{subfigure}[$\ell_0/R=1$] 
{
\begin{minipage}{0.30\textwidth}
\centering
\hspace{-0.3cm}
\includegraphics[width=0.9\textwidth]{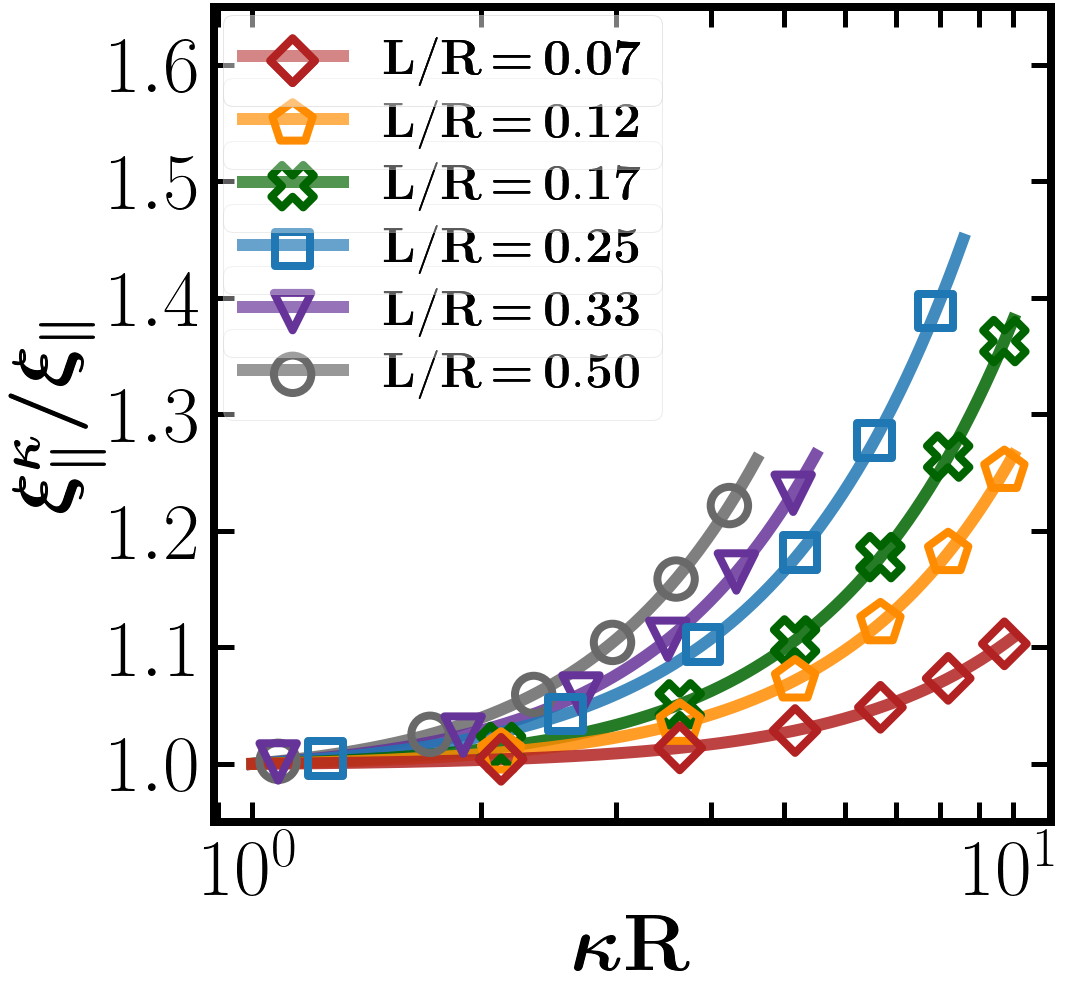}
\label{fig:pa_off_1}
\end{minipage}
}
\end{subfigure}
\begin{subfigure}[$\ell_0/R=100$ ] 
{
\begin{minipage}{0.30\textwidth}
\centering
\hspace{-0.6cm}
\includegraphics[width=0.9\textwidth]{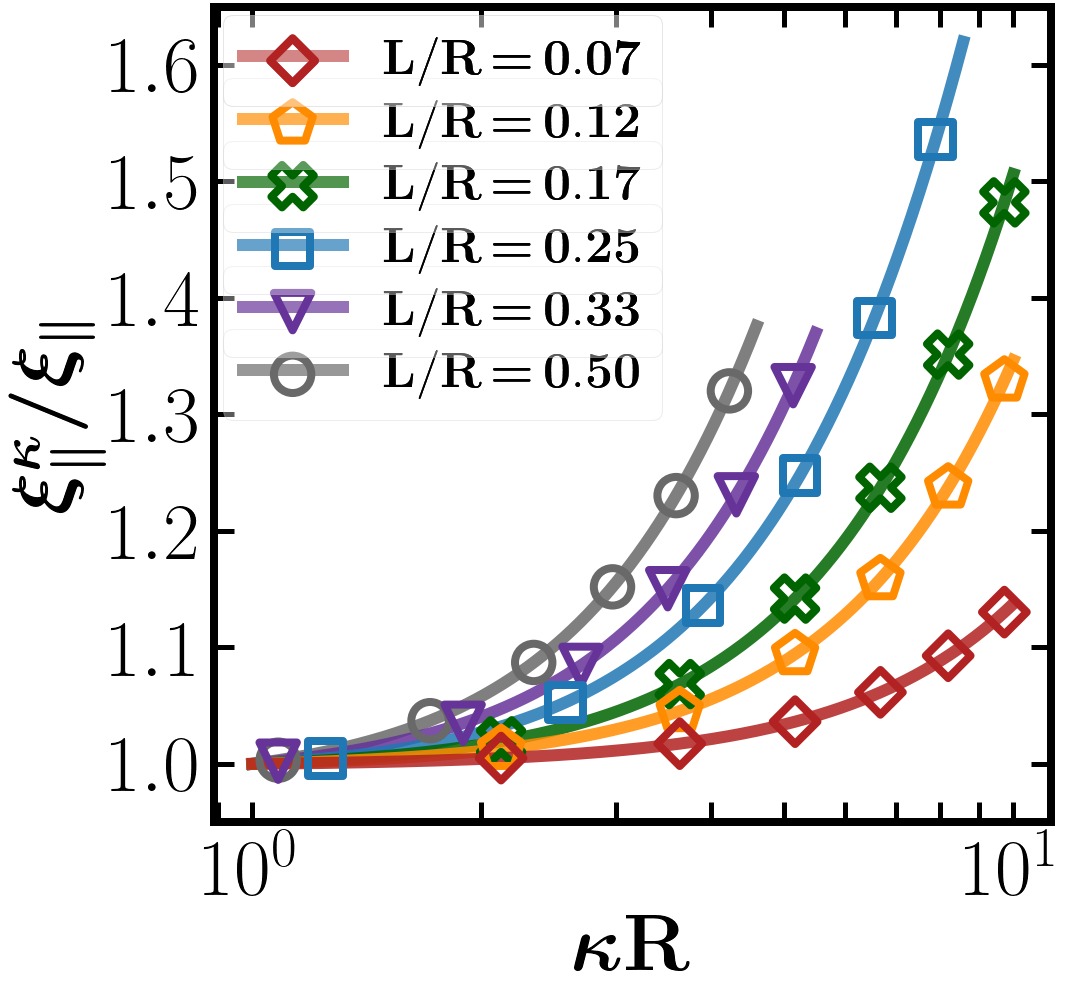}
\label{fig:pa_off_100}
\end{minipage}
}
\end{subfigure}
\begin{subfigure}[$\ell_0/R=0.01$ ] 
{
\begin{minipage}{0.3\textwidth}
\centering
%\hspace{-0.2cm}
\includegraphics[width=0.9\textwidth]{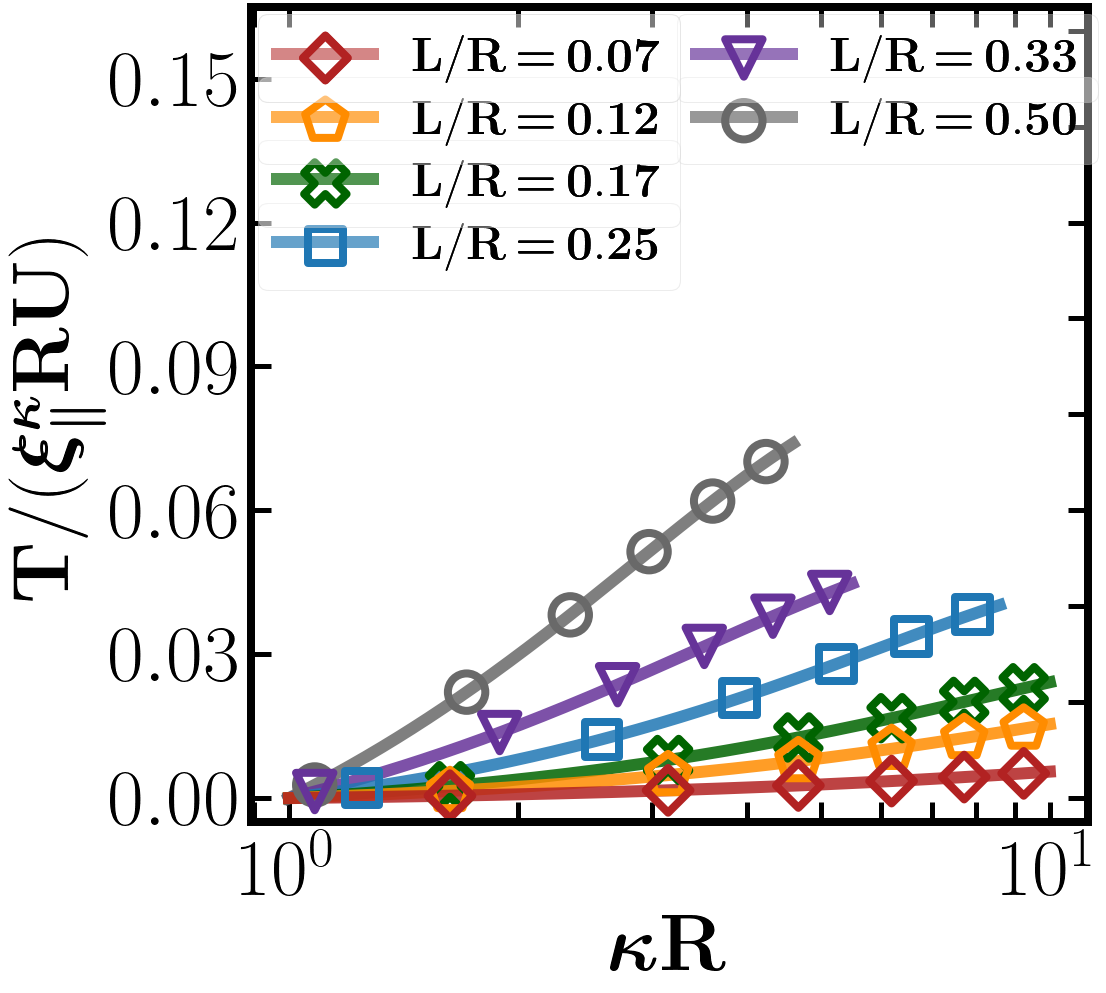}
\label{fig:pa_off_001_T}
\end{minipage}
}
\end{subfigure}
\begin{subfigure}[$\ell_0/R=1$] 
{
\begin{minipage}{0.3\textwidth}
\centering
%\hspace{-0.45cm}
\includegraphics[width=0.9\textwidth]{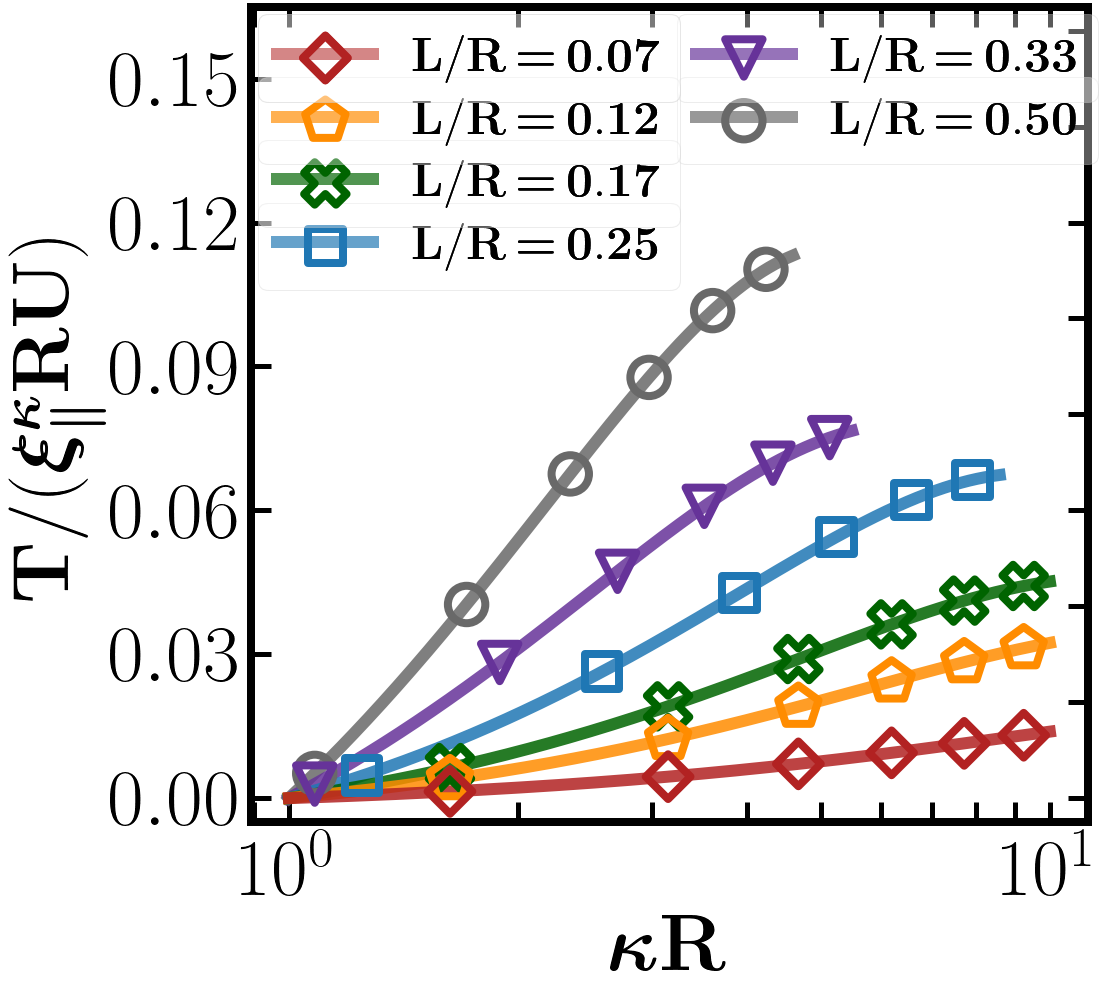}
\label{fig:pa_off_1_T}
\end{minipage}
}
\end{subfigure}
\begin{subfigure}[$\ell_0/R=100$] 
{
\begin{minipage}{0.3\textwidth}
\centering
%\hspace{-0.4cm}
\includegraphics[width=0.9\textwidth]{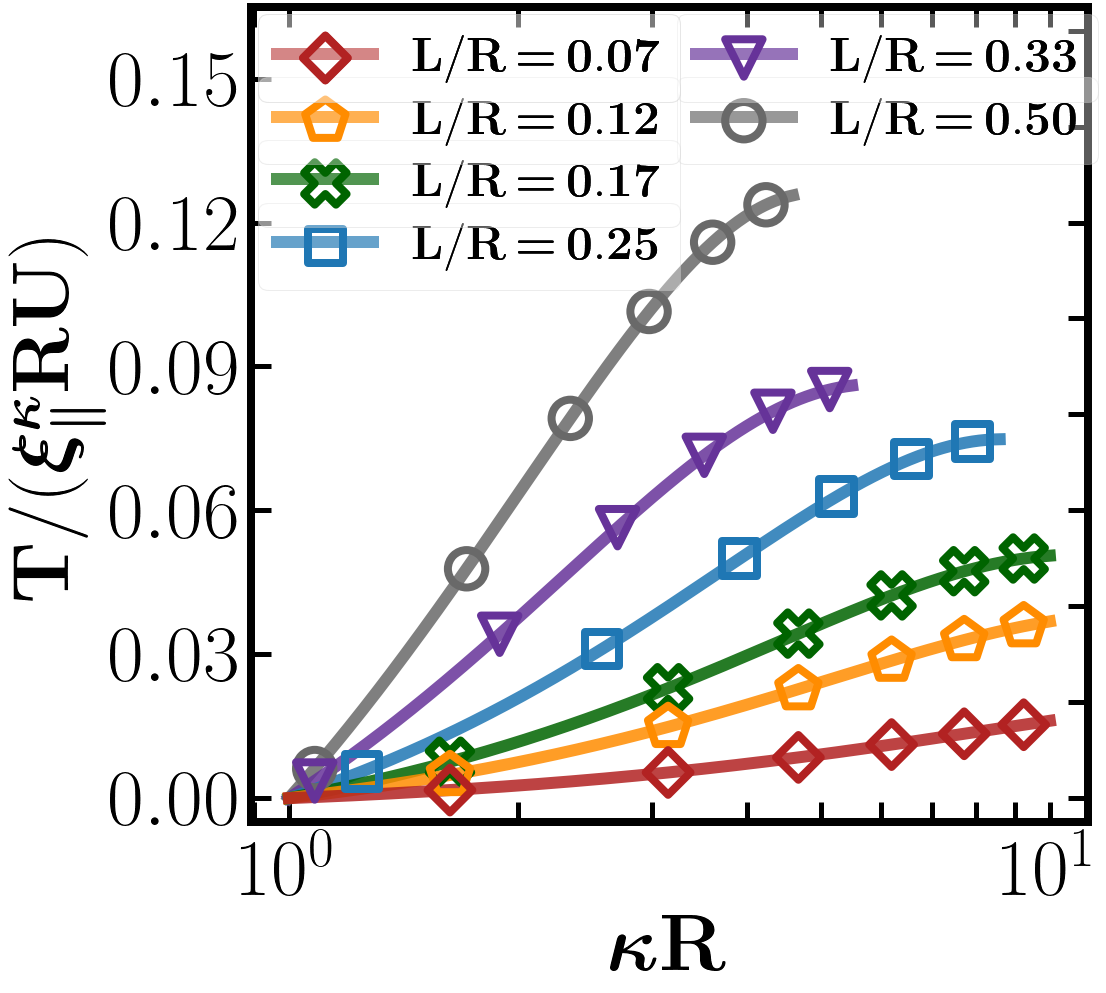}
\label{fig:pa_off_100_T}
\end{minipage}
}
\end{subfigure}
\caption{Top row (a-c): The resistance along the parallel direction vs the filament's dimensionless curvature ($\kappa R$) for $0.07\le L/R\le 0.50$ at (a) $\ell_0/R=0.01$, (b) $\ell_0/R=1$ and (c) $\ell_0/R=100$. When $\ell_0/R\ll 1$ the parallel resistance remains nearly constant with curvature and equal to the results at the equator. When $\ell_0/R \ge 1$, the resistance monotonically increases with curvature, and the rate of this increase is amplified with increasing $L/R$. 
Bottom row (d-f): the net torque induced on the filament by its motion in the parallel direction as a function of the filament's dimensionless curvature for the range $0.07 \le L/R \le 0.50$ at (d) $\ell_0/R=0.01$, (e) $\ell_0/R=1$ and (f) $\ell_0/R=100$. The torque magnitude is non-dimensionalized by the external force moment on the membrane: $\xi^{\kappa}_\parallel R U$. Regardless of the choice of $\ell_0/R$, the torque monotonically increases with $\kappa R$; this effect is amplified with increasing $L/R$. The results appear to be a weak function of $\ell_0/R$ and the torque remains $<15\%$ of the external force moment. 
}
\label{fig:resist_off}
\end{figure}
\par
\cref{fig:pa_off_001,,fig:pa_off_1,,fig:pa_off_100}, show the variations of parallel drag of a filament with curvature $\kappa$ to its parallel drag at the equator ($\xi_\parallel^\kappa/\xi_\parallel$) as a function of the filament's curvature, at different ratios of $L/R$ and $\ell_0/R=0.01,\, 1$ and $100$. As shown earlier, when $\ell_0/R\ll 1$, the filament dynamics becomes independent of the membrane geometry, resulting in small ($<10\,\%$) increases in parallel drag with curvature, as shown in \cref{fig:pa_off_001}. 
\par
At $\ell_0/R=100$, the parallel drag shows significantly larger increase with the filament's curvature (up to $50\,\%$ increase), which can be explained as follows.
The confinement flows are controlled by the membrane geometry, when $\ell_0/R \ge  1$. Larger curvatures represent larger displacements from the equator, corresponding to stronger flow confinements on the filament's side close to the poles, leading to larger drags. These flow confinement effects increase with increasing $L/R$, resulting in larger positive deviations from the filament's drag on the equator. Nevertheless, these positive deviations remain of $\mathcal{O}(1)$. 
This is consistent with our earlier observation that 
parallel motion is only weakly affected by the flow confinement effects. 
As shown in \cref{fig:pa_1}, the choice of $\ell_0/R=1$ gives nearly the same results as $\ell_0/R=100$, which shows the fluid flows are already dominated by membrane geometry in this limit.  
%When the filament is placed off the equator ($\kappa R>1$), here we use $1/\sin\theta$ which equals to 1 on the equator and monotonically increase as filament moving up towards north pole. As discussed in \cref{sec:resist}, again, we separate the resistance behavior into three regimes: $\ell_0/R \ll 1$, $\ell_0/R \gg 1$ and $\ell_0/R \sim\mathcal{O}(1)$. In all the scenarios, the ratio of $\ell_0/R$ and $R$ are fixed while varying the length of the filament and its latitude on the sphere.
%As a results of this break of symmetry, the off-equator parallel motion produces a net torque on the filament. As shown in \cref{fig:pa_off_001}, the resistance in parallel direction of off-equator, normalized by the on-equator resistance, in the limit of $\ell_{0}/R\ll 1$, although very gently, increases with latitude and $L/R$. This result indicates that resistance in parallel direction is insensitive to the latitude which is consistence to our previous statement that parallel direction is subtly affected by the confinement effect.
%\par
%While in the opposite limit of $\ell_{0}/R\gg 1$, shown in \cref{fig:pa_off_100}, the resistance, again, monotonically increase with latitude and $L/R$. Meanwhile, the resistance increases faster than $\ell_{0}/R\ll 1$, but still less than $20\%$. The \cref{fig:pa_off_1} shows the case of $\ell_{0}/R\sim\mathcal{O}(1)$ which follows the same law as stated above.
\par
\cref{fig:pa_off_001_T,,fig:pa_off_1_T,,fig:pa_off_100_T} show the torque induced by the flow asymmetry vs the filament's curvature at $\ell_0/R=0.01,\,1$ and $100$, respectively. 
The torque magnitudes, $T$, are non-dimensionalized by their associated drag times membrane radius, $\xi_\parallel^\kappa U R$, which corresponds to the external torque on the system.
In all cases the torque monotonically increases with increasing curvature ($\kappa R$) and $L/R$. This is to be expected, as the increase in both the filament's length and its curvature leads to stronger flow asymmetries. 
%After the investigation of the resistance in parallel direction, we turn to the the magnitude of the torque induced by the parallel motion. Here, we demonstrate the dimensionless torque given a unit parallel motion of filament off-equator shown in \cref{fig:pa_off_001_T}, \cref{fig:pa_off_100_T} and \cref{fig:pa_off_1_T} corresponding to the case $\ell_{0}/R\ll 1$, $\ell_{0}/R\gg 1$ and $\ell_{0}/R \sim\mathcal{O}(1)$, respectively. The induced torque, from zero on the equator, monotonically increases as a function of latitude, $L/R$ and $\ell_{0}/R$ which shows that the break-symmetry effect is amplified on the higher latitude and smaller sphere. 
%
\section{Concluding remarks}
\noindent 
The transport and dynamic organization of biopolymers bound to the cell membrane occur in many cellular processes. Most studies on the hydrodynamics of inclusions in fluid interfaces have been on 
planar membranes, and those studies that explore the effect of membrane geometry/curvature have not considered rod-like particles or filaments \citep{Henle2008, sigurdsson2016hydrodynamic, samanta2021vortex}. 
This work fills some of the gap in the literature by computing the resistance/mobility of a single filament of constant curvature moving in a fluid spherical membrane, using a slender-body formulation. 
Our results show that the membrane spherical geometry influences the filament's dynamics in three distinct ways:
\begin{enumerate}
\item As a consequence of the boundedness of spherical geometry, the momentum is transported from the interface to the 3D fluid domains over the length-scale of the membrane's radius, when the radius is smaller than SD length. This behavior has also been discussed in \citet{Henle2010}.
When the momentum transfer length is defined as the half of the harmonic average of SD length and membrane radius, $\ell^\star=(\ell_0^{-1}+R^{-1})^{-1}$, 
the computed parallel and rotational resistance values over a wide range of $\ell_0/R$ and $L/R$ collapsed to the results of a planar membrane; see \cref{fig:pa_001,fig:pa_100,fig:pa_1,fig:ro_001,fig:ro_100,fig:ro_1}. 

\item We found that the combination of surface flow incompressibility and boundedness of spherical geometry gives rise to strong flow confinement effects (see \cref{fig:flow_pe_1_irr}) that lead to large increases in the filament's perpendicular drag. These effects increase in magnitude with increasing $L/R$; see \cref{fig:pe_001,,fig:pe_100,,fig:pe_1}. 
Related to this, we found that at $L/R>1$, the resistance diverges with the inverse of gap size: $\xi_\perp \sim (\pi R/L-1)^{-1}$; see \cref{fig:MobilityGap}. These effects cannot be mapped to planar membrane results by replacing $\ell_0$ with $\ell^\star$. 

\item As a model for curvature sensors with a preferred curvature, such as BAR-domains, we studied the case where the filament's curvature is constant along its length but larger than its minimum value $\kappa> R^{-1}$. 
We showed that this curvature mismatch generates asymmetric flows around the filament when it is translating along its axis  (see \cref{fig:flow_pa_1_off}), resulting in a net torque on it. This coupling between the parallel and rotational motions of the filament is amplified with increasing the filament's curvature and length; see \cref{fig:pa_off_001_T,,fig:pa_off_1_T,,fig:pa_off_100_T}. 
\end{enumerate}

\par
A standard method for characterizing the rheology of the membrane and other fluid interfaces is to use their measured diffusion and the predicted values of mobility
to extract the interface viscosity, $\eta_m$ (or its linear viscoelastic response, if generalized Stokes-Einstein relationship is used). When the particle is significantly smaller than SD length, the mobility (and diffusion) approaches Saffman's limit: $D =k_b T (1/4\pi \eta_m) \left[\ln (2\ell_0/a)-\gamma\right]$.
Hence, large variations of particle/probe size and viscosity produces only weak variations in the diffusive motion. In the other limit of having a filament much larger than SD length moving in a planar membrane, the dynamics is dominated by the surrounding fluid and weakly dependent on membrane viscosity. This makes rheological characterization of interfaces very challenging. 
%and the results can change greatly depending on the details of mechanical coupling between the probe, the interface and the surrounding fluid domains \citep{gambin2006lateral, Naji2007, sigurdsson2013hybrid, Camley2015}. 
    
Using the filaments' motion on a spherical geometry may resolve some of these complications. As shown in \cref{fig:pa_100}, when the membrane radius is much smaller than SD length, $\ell_0/R \gg 1$, its dimensionless parallel drag ($\xi_\parallel/(4\pi \eta_{\text{m}})$) is independent of $\ell_0$ and only a function of $L/R$. Note that, as long as $\ell_0/R\gg 1$, the drag (including the rotational and perpendicular ones) is linearly dependent on $\eta_m$. This behavior is fundamentally different from the drag on planar membranes at $L/\ell_0 \gg 1$, which is weakly (logarithmic) dependent on $\eta_m$.  Also, the dependency of $\xi_\parallel$ to $L/R$ is significantly stronger than logarithmic dependency of particles on planar membranes in Saffman limit. Hence, varying the filament's length or the membrane radius is expected to provide larger changes in the diffusion (mobility) of the particles and more accurate  measurements of $\eta_{\text{m}}$, compared to the experiments on planar membranes. Assuming membrane viscosity varies within the range $\eta_m \in [10^{-9}, \,10^{-6}]$ (Pa$\,\cdot\,$s$\,\cdot\,$m) \citep{sakuma2020viscosity},  we have $\ell_0 \in [10^{-6},\,10^{-3}]$ (m). Using vesicles/cells of a few micron in radius is likely to satisfy the condition of $\ell_0/R\gg 1$. 

Surface incompressibility and the coupling of interfacial flows to the surrounding 3D fluid domains introduce several qualitative changes in HIs between the bound particles compared to particles in 3D fluids. These changes can result in novel emergent behaviors in the interface-bound suspensions \citep{oppenheimer2019rotating, manikantan2020tunable, samanta2021vortex}. 
Our work highlights one of these changes, namely the large difference between the mobility of filaments in the perpendicular and parallel directions, when the filament's length is comparable to the membrane radius.
Varying the ratio $\chi_\perp/\chi_\parallel$ is expected to lead to qualitatively different organization of the filaments. Computational studies of suspensions of bound filaments are needed to explore some of these effects. The formulation and the results presented here form the basis for performing these simulations. 

%\newpage
%\acknowledgements
\begin{acknowledgments}
 %We thank H. Manikantan and N. Oppenheimer for their insightful comments and careful reading of the article. 
 We acknowledge support by the National Science Foundation under Career Grant No. CBET-1944156. 
\end{acknowledgments}

\appendix
\section{}\label{sec:appA}
Here we outline the fundamental solutions to the system of equations \ref{eq:Eqs}, in response to a point-force on the membrane at $(\theta_0,\phi_0)$ on the sphere, where $\theta \in [0,\pi]$ is the polar angle and $\phi \in [0,2\pi)$ is the azimuthal angle, defined in \cref{fig:illustration}. The equations after including the externally applied point-force are:
\begin{subequations}
\begin{align}
 &\eta^{\pm}\nabla^{2}\mathbf{u}^{\pm}-\nabla p^{\pm}=\mathbf{0},&
\nabla\cdot\mathbf{u}^{\pm}=0,&\\
&\eta_{m}\left(\Delta_\gamma \mathbf{u}_{m}
+K(\mathbf{x}_m)\mathbf{u}_{m}\right)-\nabla_\gamma p_{m} +\mathbf{T}|_{r=R}+\mathbf{f}^\text{ext}{\delta} (\theta_0,\phi_{0})=\mathbf{0}, &\nabla_\gamma\cdot \mathbf{u}_m=0,&
\end{align}
\label{eq:AppEqs}
\end{subequations}
where $\delta(\theta_0,\phi_0)$ are Dirac delta function. 
The analytical solutions to \cref{eq:AppEqs} was provided by \cite{Henle2008,Henle2010}, which we reproduce here for completeness. The velocity field at an arbitrary point $(\theta,\phi)$ is 
$\mathbf{u}(\theta,\phi)=\mathbf{G}(\theta,\phi,\theta_0,\phi_0)\cdot \mathbf{f} (\theta_0,\phi_0)$. Writing this expression in matrix form gives:
\[
\begin{bmatrix}
    {u}_{\theta} \\
    {u}_{\phi} \\
\end{bmatrix}
=
\frac{1}{4\pi\eta_{m}}
\begin{bmatrix}
G_{\theta\theta} & G_{\theta\phi} \\
G_{\phi\theta} & G_{\phi\phi} \\
\end{bmatrix}
\cdot
\begin{bmatrix}
    {f}_{\theta}^\text{ext}(\theta_0,\phi_0) \\
    {f}_{\phi}^\text{ext} (\theta_0,\phi_0) \\
\end{bmatrix},
\]
where 
\begin{subequations}
\begin{align}
     G_{\theta\theta}=\sum_{l=2}^{\infty} &\frac{2l+1}{s_{l}l(l+1)}
    \Big{(} 
    -P^{2}_{l}(\cos{\psi})\sin^{-2}{\psi}\sin{\theta}\sin{\theta_{0}}\sin^{2}(\phi-\phi_{0}) \\
    \nonumber
    &-P^{1}_{l}(\cos{\psi})\sin^{-1}{\psi}\cos{(\phi-\phi_{0})}
    \Big{)}\\
    G_{\theta\phi} = \sum_{l=2}^{\infty}& \frac{2l+1}{s_{l}l(l+1)}
    \Big{(} P^{2}_{l}(\cos{\psi})\sin^{-2}{\psi}
    (-\cos{\theta}\sin{\theta_{0}}+\sin{\theta}\cos{\theta_{0}}\cos{(\phi-\phi_{0})}) \\
    \nonumber 
    &\cdot
    \sin{\theta_{0}}\sin{(\phi-\phi_{0})} -P^{1}_{l}(\cos{\psi})\sin^{-1}{\psi}\cos{\theta_{0}}\sin{(\phi-\phi_{0})}
    \Big{)}\\
    G_{\phi\theta} =\sum_{l=2}^{\infty}& \frac{2l+1}{s_{l}l(l+1)}
    \Big{(}
    P^{2}_{l}(\cos{\psi})\sin^{-2}{\psi}
    (\sin{\theta}\cos{\theta_{0}}-\cos{\theta}\sin{\theta_{0}}\cos{(\phi-\phi_{0})}) \\
    \nonumber
    &\sin{\theta}\sin{(\phi-\phi_{0})}
    +P^{1}_{l}(\cos{\psi})\sin^{-1}{\psi}\cos{\theta}\sin{(\phi-\phi_{0})}
    \Big{)},\\
    G_{\phi\phi} = \sum_{l=2}^{\infty}& \frac{2l+1}{s_{l}l(l+1)}
    \Big{(}
    P^{2}_{l}(\cos{\psi})\sin^{-2}{\psi}
    (-\cos{\theta}\sin{\theta_{0}}+\sin{\theta}\cos{\theta_{0}\cos{(\phi-\phi_{0})}}) \\
    \nonumber 
    &\cdot(-\sin{\theta}\cos{\theta_{0}}+\cos{\theta}\sin{\theta_{0}\cos{(\phi-\phi_{0})}}) \\
    \nonumber 
    &-P^{1}_{l}(\cos{\psi})\sin^{-1}{\psi}
    (\sin{\theta}\sin{\theta_{0}}+\cos{\theta}\cos{\theta_{0}\cos{(\phi-\phi_{0})}})
    \Big{)},
    \end{align}
\end{subequations}
\begin{equation}
    \cos{\psi}=\cos{\theta}\cos{\theta_{0}}+\sin{\theta}\sin{\theta_{0}}\cos{(\phi-\phi_{0})},
\end{equation}
and
\begin{equation}
s_{l}=l(l+1)-2+\frac{R}{\ell_{-}}(l-1)+\frac{R}{\ell_{+}}(l+2).
\end{equation}
Here, $P_{l}^{m}(\cos{\psi})$ is the Associated Legendre polynomials with degree $l$ and order $m$, $\ell_{\pm}=\eta_{m}/\eta^{\pm}$ and $R$ is the radius of the sphere. Note that the summation of $l$ in the Green's function starts from $l=2$ where we exclude the rigid-body rotation term $l=1$ because we only consider \emph{relative} motion of filament with respect to the spherical membrane \citep{Henle2010, samanta2021vortex}.

\typeout{} 
\bibliography{bibliography}

\end{document}